\documentclass[twocolumn,twocolappendix]{aastex631}
\usepackage{amsmath}

\begin{document}

\title{Deep Optical Follow-up Observations to IceCube Cosmic Neutrinos:\\
a case for IC230724A with Subaru/HSC and prospects with Rubin/LSST}

\author[0000-0003-2579-7266]{Shigeo S. Kimura}
\email{shigeo@astr.tohoku.ac.jp}
\affiliation{Frontier Research Institute for Interdisciplinary Sciences, Tohoku University, Sendai, Miyagi 980-8578, Japan}
\affiliation{Astronomical Institute, Graduate School of Science, Tohoku University, Sendai, Miyagi 980-8578, Japan}

\author[0000-0001-8253-6850]{Masaomi Tanaka}
\affiliation{Astronomical Institute, Graduate School of Science, Tohoku University, Sendai, Miyagi 980-8578, Japan}
\affiliation{Division for the Establishment of Frontier Sciences, Organization for Advanced Studies, Tohoku University, Sendai, Miyagi 980-8577, Japan}

\author[0009-0002-5156-7819]{Seiji Toshikage}
\affiliation{Astronomical Institute, Graduate School of Science, Tohoku University, Sendai, Miyagi 980-8578, Japan}

\author[0000-0001-7449-4814]{Tomoki Morokuma}
\affiliation{Astronomy Research Research Center, Chiba Institute of Technology, Chiba, Chiba 275-0016, Japan}

\author[0000-0001-6857-1772]{Nobuhiro Shimizu}
\affiliation{International Center for Hadron Astrophysics, Chiba University, Chiba, Chiba 263-8522, Japan}

\author[0000-0001-8537-3153]{Nozomu Tominaga}
\affiliation{Division of Science, National Astronomical Observatory of Japan, National Institutes of Natural Sciences, Mitaka, Tokyo 181-8588, Japan}
\affiliation{Department of Astronomical Science, School of Physical Sciences, The Graduate University for Advanced Studies (SOKENDAI), Mitaka, Tokyo 181-8588, Japan}
\affiliation{Department of Physics, Faculty of Science and Engineering, Konan University, Kobe, Hyogo 658-8501, Japan}

\author{Naoki Yasuda}
\affiliation{Kavli Institute for the Physics and Mathematics of the Universe (WPI), The University of
Tokyo Institutes for Advanced Study, The University of Tokyo, Kashiwa, Chiba 277-8583, Japan}

\author[0000-0001-6161-8988]{Yousuke Utsumi}
\affiliation{NAOJ Chile, National Astronomical Observatory of Japan, National Institutes of Natural Sciences, Vitacura, Santiago Chile}
\affiliation{Advanced Technology Center, National Astronomical Observatory of Japan, National Institutes of Natural Sciences, Mitaka, Tokyo 181-8588, Japan}
\affiliation{Vera C. Rubin Observatory, La Serena, 1720236, Chile}

\author[0000-0002-9948-1646]{Michitoshi Yoshida}
\affiliation{Hiroshima Astrophysical Science Center, Hiroshima University, Higashi-Hiroshima, Hiroshima 739-8526, Japan}
\affiliation{Subaru Telescope, National Astronomical Observatory of Japan,  National Institutes of Natural Sciences, Mitaka, Tokyo 181-8588, Japan}
\affiliation{Department of Astronomical Science, School of Physical Sciences, The Graduate University for Advanced Studies (SOKENDAI), Mitaka, Tokyo 181-8588, Japan}

\author[0000-0002-0921-8837]{Yasushi Fukazawa}
\affiliation{Department of Physical Sciences, Hiroshima University, Higashi-Hiroshima, Hiroshima 739-8526, Japan}

\author[0000-0001-6099-9539]{Koji S. Kawabata}
\affiliation{Hiroshima Astrophysical Science Center, Hiroshima University, Higashi-Hiroshima, Hiroshima 739-8526, Japan}



\begin{abstract}
IceCube has been detecting cosmic high-energy neutrinos for more than 10 years, but their major sources are still under debate. To identify them, IceCube is issuing neutrino alerts, which enable us to perform electromagnetic follow-up observations. In this paper, we present our Subaru/HSC deep optical follow-up observations down to 25.5 mag to a well-localized neutrino event, IceCube 230724A. We conduct a dedicated analysis with extensive evaluation of background rates and true positive rates adopting the blind analysis policy to identify or disfavor tidal disruption events (TDEs) as cosmic neutrino sources. Our analysis found no TDE candidate in the region of interest. Rubin/LSST survey will enable us to constrain their fractional contribution to the cosmic high-energy neutrino background, either $\lesssim 60\%$ or $\gtrsim30\%$ for non-detection and detection, respectively, if Rubin covers the error regions of 10 neutrino events.
\end{abstract}



\section{Introduction} \label{sec:intro}

Cosmic high-energy neutrino detection was first reported by IceCube in 2013 \citep{IceCube2013PRL,IceCube2013Sci}, which opened up a new window of astrophysics. IceCube Collaboration has been detecting cosmic neutrinos for more than 10 years and reported a few associations of cosmic neutrinos with astrophysical objects, including a nearby Seyfert Galaxy, NGC 1068 \citep{IceCube2022Sci}, and Galactic plane \citep{2023Sci...380.1338I}. However, the sources of the bulk of the observed neutrinos are still elusive and have been actively discussed for more than 10 years \citep[][]{2023ecnp.book..107H,2023ecnp.book..433K,2023ecnp.book..483M}. 

Some classes of astrophysical transients, including tidal disruption events (TDEs: \citealt{hayasaki2019ApJ,Murase:2020lnu,Liu2020PhRvD}), interaction-powered supernovae \citep[e.g.][]{Zirakashvili16a,murase2018PhRvD,kimura2025apj}, and engine-driven supernovae \citep[e.g.,][]{MuraseIoka13a,Mukhopadhyay2026arXiv} are proposed as the source of cosmic neutrinos. To identify these transients as cosmic neutrino sources, IceCube developed an alert system which enables observers to perform follow-up observations to cosmic neutrino events \citep{IceCube:2016lmt}. This alert system leads to possible association between neutrino events and astrophysical transients, including a flaring blazar, TXS 0506+056 \citep{IceCube2018blazar1,IceCube2018blazar2}. In addition, supernovae and optical transient searches to IceCube alerts are extensively performed \citep{2015ApJ...811...52A,2019A&A...626A.117P,2019ApJ...883..125M,2026A&A...708A.223G}, and non-detection of plausible counterparts disfavors luminous transients as the source of cosmic neutrinos \citep{2022MNRAS.516.2455N,2023MNRAS.521.5046S,2025ApJ...993...23T}.

Several associations between neutrino events and tidal disruption events (TDEs) were also reported \citep{2021NatAs...5..510S,2022PhRvL.128v1101R,2023ApJ...953L..12J,2024MNRAS.529.2559V,2024ApJ...969..136Y,2025ApJ...990...18L}\footnote{The latest IceCube alert stream (IceCat-2) resulted in some shifts of the neutrino localization regions, which cause the first three neutrino-associated TDEs (AT2019dsg, AT2019fdr, AT2019aalc) to be located outside the neutrino error regions \citep{icecat2}.}. However, individual papers analyzed the data independently, leading to an unclear situation how significant the associations are. Also, the fractional contribution of TDEs to the observed neutrinos is still uncertain. 
Considering the source distribution in the Universe, majority of neutrino-emitting astrophysical transients should be located at redshift of $z\sim0.2-1$ and their expected apparent magnitude would be $20-25$ mag, depending on the source luminosity. Then, current follow-up observations using 1-2 m-size telescopes will be unable to detect considerable fraction of distant transients. 
Since typical error regions of IceCube neutrino events are 1-3 deg$^2$, we need wide and deep follow-up observations to identify neutrino-emitting transients.

In this paper, we present our optical follow-up observations to a neutrino event, IC230724A, using Subaru/Hyper Suprime-Cam \citep[HSC;][]{miyazaki06,HSC2018M}.
IC230724A is a well-localized GOLD neutrino event occurring at UT 01:49:13.381 on July 24, 2023, whose 90\% localization was $\lesssim0.5$ deg when the alert was issued \citep{ic23gcn}\footnote{As shown in Section \ref{sec:obs}, the localization was changed in the updated IceCat-2 preliminary data release, the data of which is available at  \href{https://dataverse.harvard.edu/dataset.xhtml?persistentId=doi:10.7910/DVN/RX28YT}{IceCat-2 preliminary data release}.}. We take advantage of Subaru/HSC's wide field-of-view and great sensitivity to look for distant transients. 
Our follow-up criteria are (i) astrophysical probability higher than 0.5, (ii) the localization angular error smaller than 0.6 deg, and (iii) continuously observable by Subaru for a few months after the alert is issued. While our team has Subaru observation time, only IC230724A satisfied all of these criteria, and thus, we do not perform follow-up observations to any other neutrino alerts.

To exclude any bias during the analysis, we follow the blind analysis policy \citep{2005ARNPS..55..141K}. Under this policy, we first need to set search criteria and estimate the number of unrelated transients and the true positive rate (TPR) of the target source using the data outside the region of interest. After that, we open the data in the region of interest with the same criteria to obtain the result. This method enables us to exclude any biased view. 
This policy is uncommon in time-domain astronomy, and only a few adopted it \citep{2019ApJ...883..125M,2025ApJ...993...23T}. 
We also discuss future prospects for identifying tidal disruption events as neutrino production sites using a future wide and deep optical survey, which will be achieved by NSF-DOE Vera C. Rubin Observatory/Legacy Survey of Space and Time  \citep[LSST;][]{2019ApJ...873..111I}.

Our paper is organized as follows. In Section \ref{sec:obs}, we describe our observations and data reduction. In Section \ref{sec:source}, we introduce our source model, a population of TDEs. Section \ref{sec:selection} discusses the selection method of the source TDE from unrelated transients and variable objects. We write the statistical analysis method in Section \ref{sec:statistics}, and our analysis results are given in Section \ref{sec:result}. We evaluate the future prospects with Rubin/LSST in Section \ref{sec:rubin} and provide summary and implications in Section \ref{sec:summary}.
We use cosmological parameters obtained by \citet{2014A&A...571A..16P}.

\begin{figure*}[t!]
\epsscale{1.0}
\plotone{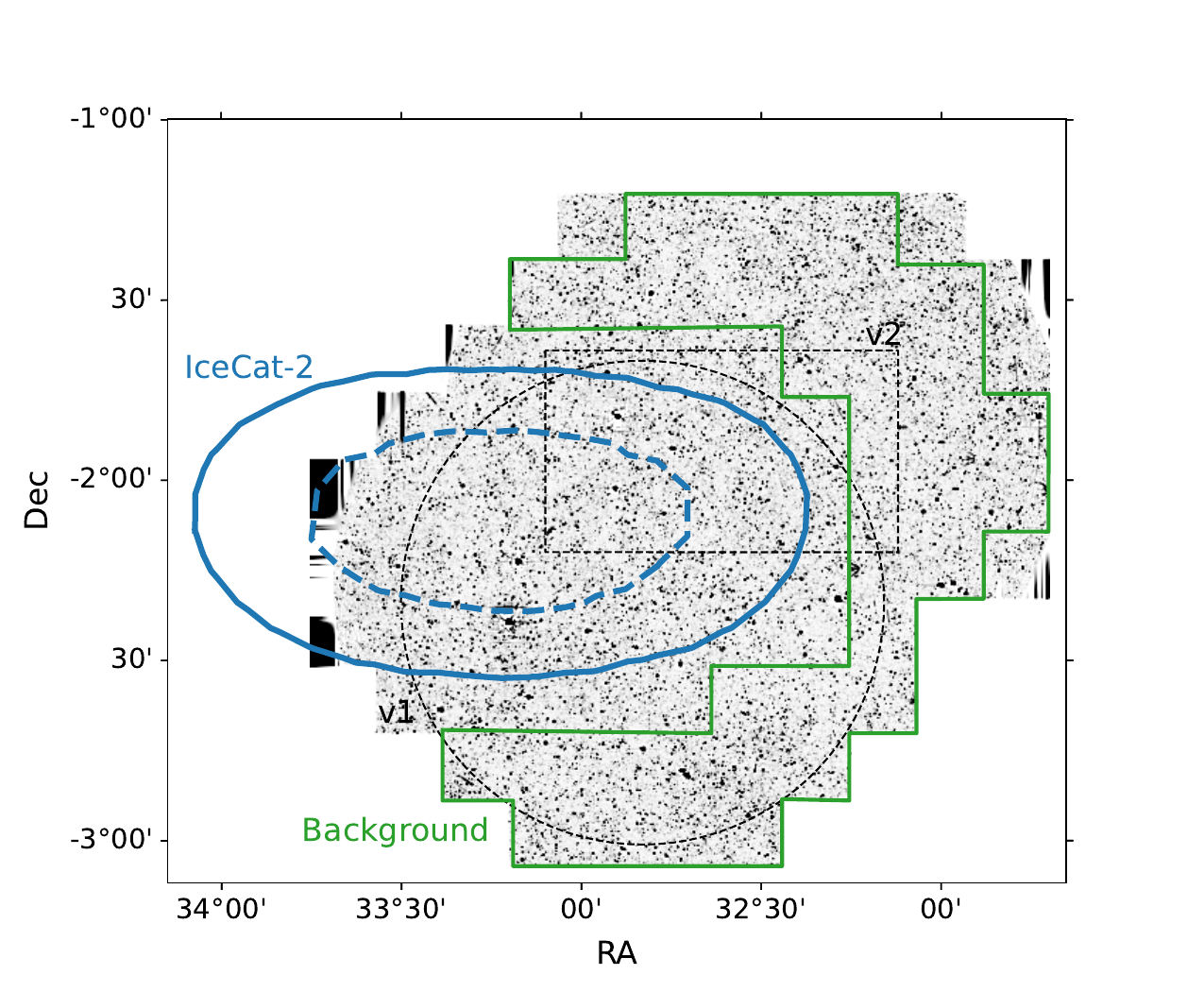}
\caption{Localization areas of IC230724A
and our HSC $r$-band image.
Black circle and rectangle show 90\% probabability region from the automatic alert (v1) and updated alert (v2), respectively.
Blue solid and dashed circles show 90\% and 50\% probability regions from IceCat-2, respectively. 
The region surrounded by green lines show our background region.
} 
\label{fig:pointing}
\end{figure*}

\begin{deluxetable}{lllll}
\tablewidth{0pt}
\tablecaption{Summary of HSC observations for IC230724A \label{tab:obs}}
\tablehead{
\colhead{Date} & \colhead{Time$^1$} & \colhead{Filter} & \colhead{FWHM} & \colhead{Depth$^2$} \\
                & (days) &        & (arcsec) & (mag)
}
\startdata
2023 Aug 10 & 18.39 & $i$ & 1.53 & 24.4 \\
            & 18.46 & $g$ & 0.99 & 25.7 \\
            & 18.52 & $r$ & 0.65 & 25.5 \\
2023 Aug 23 & 31.37 & $i$ & 0.84 & 24.8 \\
            & 31.43 & $g$ & 1.10 & 25.9 \\
            & 31.49 & $r$ & 0.87 & 25.6 \\
\enddata
\tablecomments{1. Time after the neutrino detection. 2. 5$\sigma$ limiting magnitudes.}
\end{deluxetable}

\section{Observations and Data Reduction} 
\label{sec:obs}

\subsection{Observations}

We performed Target-of-Opportunity (ToO) imaging observations
toward the localization region of IC230724A with Subaru/HSC \citep{HSC2018M}.
Our observations were conducted on 2023 August 10 and 23 UT.
We planned further follow-up observations, but it was not possible due to an issue of the prime mirror of the Subaru telescope.
The epochs of our observations correspond to 18 days and 31 days after the neutrino detection,
respectively.
Hereafter, we call these observations Epoch 1 and 2, respectively.
In each epoch, we obtained the images with three optical filters
($g$, $r$, and $i$ bands).

Images were taken toward two pointing centers (Figure \ref{fig:pointing}):
one is centered to the position estimated in the automated alert 
from the IceCube collaboration (v1),
while the other is centered to the position estimated in
the updated alert (v2, \citealt{ic23gcn}).
The field of views of two pointing positions have overlap.

For each set of filters and pointing centers,
one 30 sec image was first taken for photometric calibration.
Then, 5 set of 300 sec images were taken 
with dithering to fill the CCD chip gaps of the HSC.
The observational conditions (seeing and transparency of the sky) were excellent except for the $i$-band observations at Epoch 1,
which suffered from a poor seeing due to strong wind.
Summary of our observations is given in Table \ref{tab:obs}.

After our observations were conducted, IceCube collaboration presented an updated event catalog of alert tracks (IceCat-2, \citealt{icecat2}).
The 50\% and 90\% probability regions for IC230724A in IceCat-2 are shown with blue dashed and solid ellipses in Figure \ref{fig:pointing}, which is  centered at $(\alpha,~\delta)=(33.22^\circ,~-2.13^\circ)$.
Since this localization is estimated by an improved reconstruction technique in IceCat-2, 
we use this localization throughout the paper.
The areas for the 50\% and 90\% probability regions are 0.42 deg$^2$ and 1.18 deg$^2$, respectively.
Inside the 90\% probability region, our HSC data cover 0.918 deg$^2$. The total probability covered by our data corresponds to 73.5\%. Hereafter, we call this region our source region\footnote{We define our source region as the overlapping region of the 90\% contour and 0.7 deg from the first HSC pointing center, $(\alpha,~\delta)=(32.95^\circ,~-2.30^\circ)$.}.

The region indicated by the green line in Figure \ref{fig:pointing} 
is used for the background estimate (Section \ref{sec:selection}).
We call this region our background region.
Ideally, the background region should be large enough to reduce the statistical uncertainty in the background rate estimates.
However, to reach the sufficient sensitivity in one night, 
our observations did not cover a very wide area.
As a result, the area of our background region 
(1.44 deg$^2$) is larger than the source region only by 47\%.

\subsection{Data reduction}

The imaging data were processed in a standard manner
by using the HSC reduction pipeline ({\tt hscpipe} version 8.4 \citealt{HSC-pipe2018PASJ,ivezic08,axelrod10}).
After standard reduction for each frame
including bias, dark, and flat corrections, 
five exposure images were co-added together.
Astrometric and photometric calibrations were performed with respect to
Pan-STARRS1 catalog \citep{chambers16}.

To identify the transients, we performed image subtraction by the HSC pipeline,
which implements the algorithm developed by \citet{alard98} and \citet{alard00}.
As the localization area is included in the survey footprint of
HSC-SSP survey \citep{aihara18ssp}, we used the HSC-SSP stacked image
as the reference image.
The HSC-SSP images were taken about 2600 ($\pm 200$) days,
2800 ($\pm 20$) days, and 2200 ($\pm 700$) days before the IC230724A
for $g$, $r$, and $i$ bands, respectively.
Image subtraction algorithm searches for a space-varying convolution kernel
to match the point-spread functions (PSFs) of new and reference images.
For a better image subtraction, we performed image subtraction for each frame,
and then stacked the subtracted images.
Hereafter, these stacked subtracted images are simply denoted as subtracted images.

Finally, source detection was performed for the subtracted images.
We imposed a condition that transient candidates of interest should be detected, i.e., the subtracted flux $\ge5\sigma$, both in $g$-band and $r$-band images at Epoch 1. 
Then, for these candidates, we performed forced photometry in all the subtracted images to provide the light curves.
The detected sources include extragalactic transients such as SNe and TDEs, variable objects such as stars and active galactic nuclei (AGNs), as well as fake detection (bogus) due to the imperfect image subtraction. 
Further selection processes for these candidates are discussed in Section \ref{sec:selection}.

\section{Source model}
\label{sec:source}

We test a population of TDEs as the neutrino source. 
We construct lightcurve templates for TDEs based on the ZTF-detected TDE samples given in \cite{2023ApJ...942....9H}, which reported 30 optically selected TDEs. Their bolometric lightcurve can be modeled by a Gaussian rise and power-law decay:
\begin{equation}
L_{\rm bol}(t) =\left\{   
\begin{array}{ll}
L_{\rm pk}\exp\left(-\left(\frac{t-t_{\rm pk}}{t_{\rm rise}}\right)^2\right) & ~~~(t \leq t_{\rm pk})\\
L_{\rm pk}\left(1+\frac{t-t_{\rm pk}}{t_{\rm decay}}\right)^{-p_{\rm decay}} & ~~~(t > t_{\rm pk})
\end{array}
\right. ,
\end{equation}
where $L_{\rm pk}$, $t_{\rm rise}$, $t_{\rm decay}$, $t_{\rm pk}$, and $p_{\rm decay}$ are the peak luminosity, the rise time scale, the decay time scale, the peak time, and the decay index, respectively. We assume that these TDEs exhibit the blackbody photon spectrum, whose temperature evolution is described by 
\begin{equation}
    T(t) =\left\{   
\begin{array}{ll}
T_{\rm pk} & ~~~(t \leq t_{\rm pk})\\
T_{\rm pk}+(t-t_{\rm pk})\frac{dT}{dt} & ~~~(t > t_{\rm pk})
\end{array}
\right. ,
\end{equation}
where $T_{\rm pk}$ is the temperature at $t=t_{\rm pk}$ and we assume time derivative of the temperature, $dT/dt$, is constant in time. We can construct a lightcurve by setting 6 parameters: $L_{\rm pk}$, $t_{\rm rise}$, $t_{\rm decay}$, $p_{\rm decay}$ $T_{\rm pk}$, and $dT/dt$. Examples of our lightcurve templates are shown in Figure \ref{fig:tde_lc}.
See Appendix \ref{sec:param_list} and \ref{sec:TDE_lc} for the parameter list and our construction method for TDE lightcurve templates, respectively.

\begin{figure}[t!]
\epsscale{1.0}
\plotone{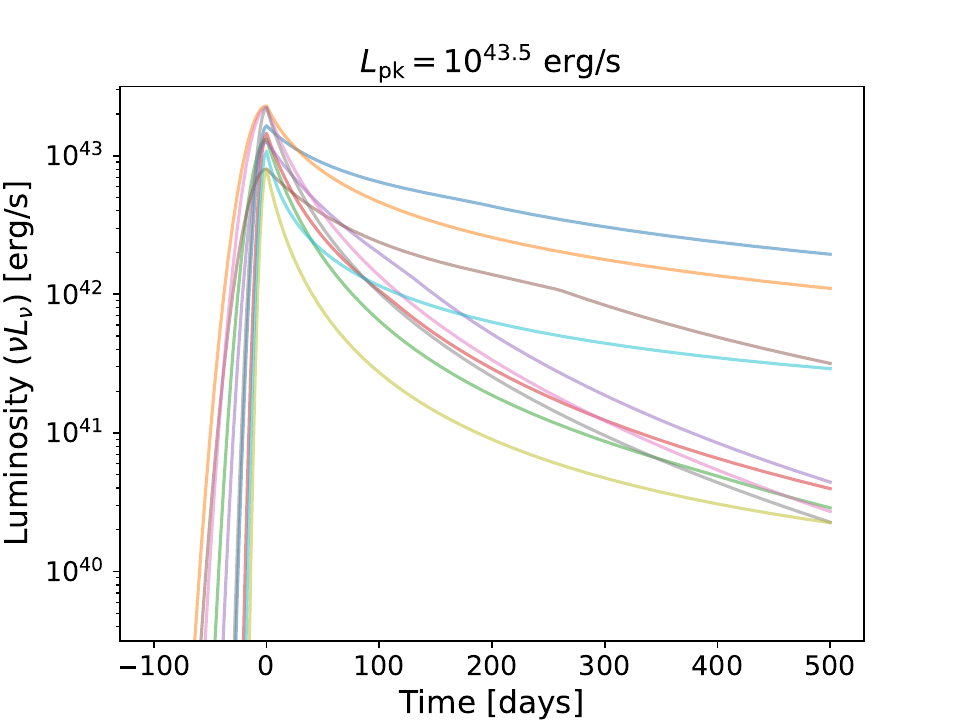}
\caption{Examples of our TDE lightcurve templates at $\nu=7.5\times10^{14}$ Hz or ($\lambda=$ 4000 \AA) at the source rest frame. We fix $L_{\rm pk}=10^{43.5} \rm~erg~s^{-1}$ and varied the other parameters to indicate diversity of the population.} 
\label{fig:tde_lc}
\end{figure}

Regarding the time lag between the neutrino event and the peak time of optical light curve, we assume that $t_{\rm pk}$ is uniformly distributed between $t=t_\nu-t_{\rm decay}$ and $t=t_\nu$. This implies that the duration of neutrino production is similar to the optical one. 
This treatment is motivated from theoretical models that produce neutrinos using outflows \citep[e.g.][]{Murase:2020lnu,2023ApJ...948...42W}.
Some scenarios argue that neutrino production can be delayed from the optical peak by more than 100 days \citep[e.g.,][]{2024ApJ...969..136Y,2024MNRAS.534.1528M}. However, optical follow-up observations in general are not suitable to constrain these scenarios, and thus, we do not discuss them.

We also take into account the luminosity function and redshift evolution of TDEs, although these are still uncertain. The neutrino counterpart should appear on the basis of the energy generation rate, and thus, we assume that the probability distribution function of the source TDE luminosity follows $L_{\rm pk}^2 (dN/dL_{\rm pk})$, where $dN/dL_{\rm pk}$ is the luminosity function of TDEs. 
We use the luminosity function in terms of the peak UV and optical blackbody luminosity for ZTF-detected TDEs given by \citet{2023ApJ...955L...6Y}.

\begin{figure}[t!]
\epsscale{1.0}
\plotone{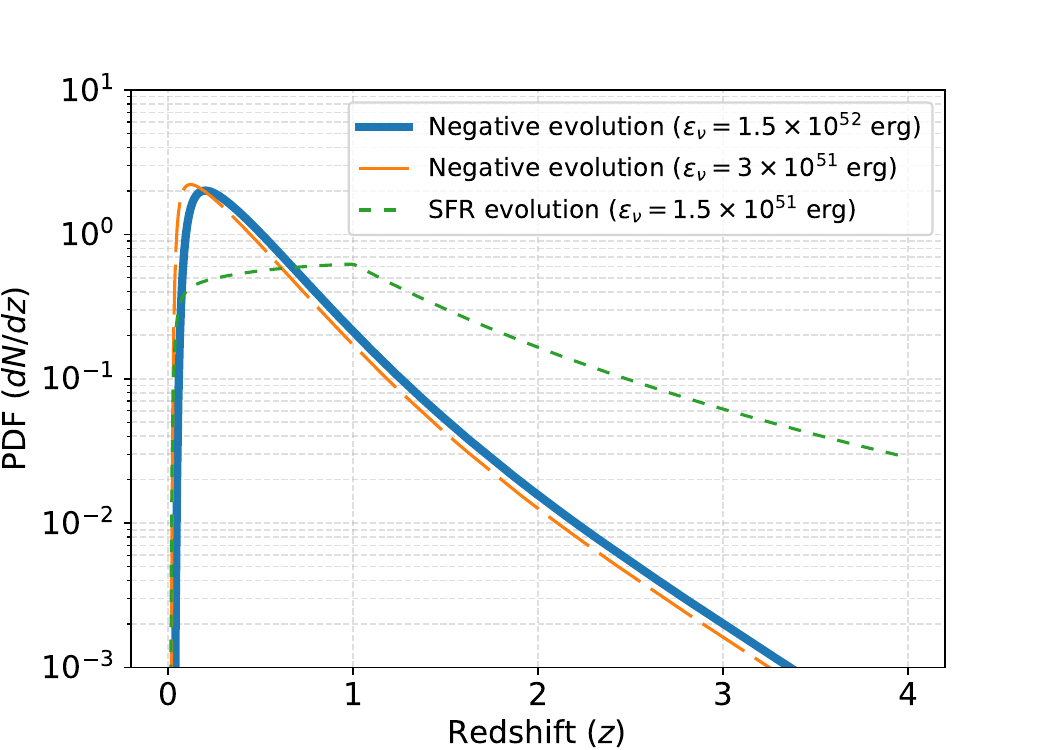}
\caption{Redshift distribution of the neutrino-emitting TDEs. The thick-solid line is our reference distribution. The thin-dashed and thin-dotted lines are for the neutrino energy fluence of $\varepsilon_\nu=3\times10^{51}$ erg and redshift evolution with the star-formation history, respectively.} 
\label{fig:zdist}
\end{figure}

The redshift distribution of the  neutrino-emitting TDEs is computed in a way similar to \cite{2025ApJ...993...23T} with some modification to compute the distribution for singlet events. First, we set the local TDE rate using the blackbody luminosity function at the break luminosity, $R_0=6\times10^{-8}\rm~Mpc^{-3}~yr^{-1}$ at $L_{\rm pk}=1.5\times10^{44}\rm~erg~s^{-1}$ \citep{2023ApJ...955L...6Y}. Then, we assume that these TDEs provide the dominant contribution to the diffuse neutrino flux detected by IceCube, which enables us to estimate the neutrino energy fluence per TDE, $\varepsilon_\nu=1.5\times10^{52}$ erg (see Eq. (2) in \citealt{2025ApJ...993...23T}). Here, we use the negative redshift evolution of TDEs given by \citet{2015ApJ...812...33S}, which is supported by IR-echo observations \citep{2025A&A...695A.228N}. Using this $\varepsilon_\nu$, the negative TDE redshift evolution, and the IceCube effective area \citep{2017APh....92...30A}, we can compute the redshift distribution of neutrino sources (see Eqs. (8) and (16) in \citealt{2022ApJ...937..108Y} by replacing the Poisson term to the singlet one). 
We cautiously note that the redshift distribution of neutrino-emitting TDEs is different from the redshift distribution of TDEs. 
The redshift distribution of the neutrino-emitting TDEs is proportional to the contribution of TDEs in each redshift bin when the expected number of neutrino signals from a TDE, $\mu_s$, is less than 1. In contrast, for a very nearby TDE happening at $z\sim0$, the neutrino fluence is so large that $\mu_s$ can be larger than 1. Then, probability of detecting multiple neutrino signals is larger than that of detecting a single neutrino. For this reason, the PDF of singlet events shown in Fig. \ref{fig:zdist} drops to 0 at $z=0$.

The resulting redshift distribution is given in Figure \ref{fig:zdist}. 
This depends on the assumed redshift distribution of TDEs and the neutrino fluence per TDE ($\varepsilon_\nu$). We examined how these two affect the resulting redshift distributions.
Since we adopt a negative redshift evolution of TDEs,
the neutrino-emitting TDEs (thick solid line) tend to locate closer than the neutrino sources with the star-formation history evolution (thin dotted line). 
On the other hand, the neutrino fluence per TDE does not significantly affect the resulting redshift distribution. 

As described in Section \ref{sec:selection}, we use the brightness contrast with respect to the host objects for the candidate selection using the light curves. Thus, we also need to estimate the flux from the host galaxies of TDEs. 
We use the $g$-band all-galaxy luminosity function given in \citet{2007ApJ...665..265F} as the luminosity function of TDE hosts. 
To estimate $g$-band flux of host galaxies, we assume that SEDs of the TDE host galaxies can be described by a power-law form of $\nu F_\nu \propto \nu^{-0.5}$. This spectrum is roughly consistent with the SEDs of blue galaxies (Sdm and Spi4) in the SWIRE spectra templates \citep{2007ApJ...663...81P}. 
In reality, the populations of TDE host galaxies are more complicated. Nevertheless, our assumptions are conservative in the sense that TDEs occuring in green-valley and elliptical galaxies are more easily identified owing to lower $g$-band luminosity. 

We consider TDEs occurring in these host galaxies and evaluate the flux from both TDEs and host galaxies in the following sections.

\section{Candidate Selection}
\label{sec:selection}

For the reduced HSC data, we apply various selection criteria to select a TDE that produces the neutrino event. We follow the blind analysis policy in our TDE search. The numbers shown in this section are obtained solely by analyzing data in our background region and simulated observations, and the results of our source region are shown in Section \ref{sec:result}. We briefly overview our selection method in Section \ref{sec:selection_overview}. Then, in Section \ref{sec:background}, we describe the estimates of background rate using the background region (1.44 deg$^2$; see Figure \ref{fig:pointing}). In Section \ref{sec:tpr}, we estimate the TPR of our selection criteria using both the background region and simulated observations.

\subsection{Overview of our selection method} \label{sec:selection_overview}

To follow the blind analysis policy, we first  estimate the number of unrelated transients (``background objects'' for our purpose) in our source region and the TPR of the source TDE.
For these purposes, we analyze the data in the background region and perform observation simulations with the source model described in Section \ref{sec:source}. 
We found that our background region in the analysis is too small to find a robust ($\sim3\sigma$) association. It is difficult to evaluate how large the background region should be to achieve robust associations in advance, but 20--30 times larger area than the source region would be desired (see Appendix \ref{sec:photometric_detail}). Although the available data is limited, we will tune our parameters such that the purity of the source is maximized (see Section \ref{sec:statistics}).   

We set three steps to select the TDE: (I) Image-quality cut where we remove fake detections caused by image subtraction,  (II) Galactic star cut where we remove Galactic objects, and (III) photometric classification where we remove non-TDE variable objects. To estimate the background rate, we adopt  data driven methods in which we solely use the observational data in the background region. To evaluate the TPR of the source TDE, we combine the data-driven and simulation-driven methods; TPR estimates for steps (I) and (II) are done with data-driven methods, whereas the TPR estimate for step (III) is done with a simulation-driven method. 
For demonstration purpose, we set the signalness of the neutrino event to 1.0. If we take the singlaness into account, TPR would be reduced by a factor of the value of signalness.

\subsubsection{Image-quality cut}

Here we describe our image-quality cut.
The image-based selection is done based on the source shape in the subtracted images to reduce the number of bogus detections.
We use full width at half maximum (FWHM)
and elongation (elon $=$ major axis/minor axis) of the detected sources
as compared with the point sources in the original image before the subtraction.
Namely, we impose the criteria of (1) 0.5 $<$ FWHM/FWHM$_{\rm ps} <$  2.0
and (2) elon/elon$_{\rm ps}$ $<$ 2.0, where FWHM$_{\rm ps}$ and elon$_{\rm ps}$ are
the FWHM and elongation of the point sources in the images before the subtraction.
\citet{ohgami21} and \citet{ohgami23} showed
that these criteria efficiently remove bogus
in their HSC follow-up observations for gravitational wave sources.
We follow the same philosophy but loosen the thresholds
to keep the high TPR of real transients.
As described in Section \ref{sec:obs}, we impose a condition that
the source should be detected and should pass these criteria
both in $g$- and $r$-band images at Epoch 1. 
Note that the spatial distribution of objects passing the image-quality cut is somewhat inhomogeneous, compared to that of sources detected in Section \ref{sec:obs}. This is because the image-quality is affected by nearby bright stars, causing a higher rejection rate around bright stars.

\subsubsection{Galactic star cut}

Transient candidates after the image-quality cut still include a large number of Galactic stars, i.e., stars with a high proper motion and variable stars.
As our reference images were taken more than 5 years before our observations,
a source with a high proper motion causes a characteristic positive-negative pattern.
Thus, if the detected source is associated with more than 10 significant negative pixels 
($F < -5 \sigma$; $F$ and $\sigma$ are the flux and 1-sigma error, respectively) within 20 $\times$ 20 pixel region
(3.36 $\times$ 3.36 arcsec, the pixel scale is 0.168 arcsec/pixel),
we judge such candidate as a high proper-motion star.

An ideal way to remove Galactic variable stars would be crossmatching with known variable star catalogs.
There is, however, no complete variable star catalogs reaching the sensitivity of our HSC images. 
Thus, in this work, to exclude only relatively bright variable stars, we crossmatch our candidates with the Gaia DR3 variable source catalog (down to $\sim 20$ mag; \citealt{2023A&A...674A..14R}).
We exclude transient candidates if their positions match within 2 arcsec from known sources in the Gaia variable source catalog. The Gaia variable source catalog also contains AGN, and we also remove transient candidates within 2 arcsec from the Gaia-cataloged AGN.

\subsubsection{Photometric classification cut}

For the remaining objects, we distinguish TDEs from other transients/variables using their lightcurve features, such as bluer colors, longer durations, and strong variability ampiltudes. First, to evaluate lightcurve features, we define the maximum and minimum fluxes in each band as $F_{j,\rm max/min} = F_j \pm \sigma_j$, where subscript $j=g,~r,$ or $i$ indicates the photometric band.
To quantitatively separate the source TDE from other variables/transients, we introduce the following 7 criteria:

\begin{enumerate}
\def\labelenumi{(\arabic{enumi})}
    \item $5\sigma$ detection at $t=t_1$:\\ Since the TDE should be brightest at the first epoch, we require detection at the first epoch in $g$ and $r$ bands, $F_g(t_1) \ge 5\sigma_g ~~{\rm and}~~ F_r(t_1) \ge 5\sigma_r$\footnote{We use aperture photometry in this section, whereas we use PSF photometry for the detection procedure discussed in Section \ref{sec:obs}.}. We do not impose this condition for $i$ band because of the poor seeing at Epoch 1.
    
    \item Long duration: \\
    SNe have usually shorter duration than TDEs.  For our Subaru observation, we have the data only at two epochs, and thus, we require 4 detection (i.e., $F_g \ge 5\sigma_g$ and $F_r \ge 5\sigma_r$) on 2 epochs (i.e., 2 filters $\times$ 2 epochs). We do not require $i$-band detection because the data quality is poor. This criterion also help removing asteroids because they usually move during our observation epochs. 

    \item Monotonically declining:\\
    We assume that neutrino detection was delayed compared to the optical peak time, which should lead to a monotonically declining lightcurve. We discard objects that satisfy $F_{j,\rm max}(t=t_1) < F_{j,\rm min}(t=t_2)$ where $j=g,~r,$ or $i$ band.  

    \item non-V-shape SED: \\
    We assume that TDEs indicate a blackbody-like spectrum. Then, the V-shape SED, i.e., $F_{g,\rm min} / F_{r,\rm max} > 1$ and $F_{i,\rm min}/F_{r,\rm max} >1$, is not achieved by a blackbody spectrum. We discard objects showing such a V-shape SED. This is often seen in AGN or variable stars in case the reference data for different bands were taken at different dates. 

    \item Significant variation among epochs:\\
    TDEs typically show a faster and stronger variability than AGN. We discard the object showing a weak variability, i.e., $F_{j,\rm max}(t=t_{\rm max})/F_{j,\rm min}(t=t_{\rm min})\equiv \mathcal{R}_{j,\rm Mm} < \mathcal{R}_{\rm Mm,th}$ for $j=g$ and $r$ bands, where $t_{\rm max} $ and $t_{\rm min}$ are the epochs of the maximum and minimum fluxes, respectively. We do not impose this criteria for $i$ band. We use the same value of $\mathcal{R}_{\rm Mm,th}$ for both bands.

    \item Blue color at all epochs: \\
    TDEs typically have higher temperatures than SNe. We use the ratio of two consecutive bands to distinguish TDEs from SNe. we discard objects that satisfy $F_{j,\rm max}/F_{l,\rm min}\equiv \mathcal{R}_{jl,\rm color} < \mathcal{R}_{\rm color,th}$ at both epochs for $(j,~l)=(g,~r)$ and $(r,~i)$. We use the same value of $\mathcal{R}_{\rm color,th}$ for both bands.

    \item High flux ratio to host galaxies: \\
    TDEs are observed as flares in non-AGN galaxies, which leads to a stronger flux variation in a bluer band than typical AGN. Also, TDEs are more luminous than SNe, making a higher contrast to their host galaxies. We compare the subtracted flux in $g$ band, $F_g(t=t_1)$, to the flux of the nearest object in the reference image (most likely it should be the host galaxy), $F_{g,\rm host}$. We discard the objects that satisfy $F_g(t=t_1)/F_{g,\rm host} \equiv \mathcal{R}_{g,\rm host} < R_{g,\rm host,th}$. 
\end{enumerate}

This photometric classification scheme has 3 parameters ($\mathcal{R}_{\rm Mm,th}$, $\mathcal{R}_{\rm color,th}$, $\mathcal{R}_{g,\rm host,th}$). We tune these parameters to maximize the ratio of the TPR to the background rate (see Section \ref{sec:statistics}).

\subsection{Estimates of background rates}
\label{sec:background}

We apply our (I) image-quality cut, (II) Galactic star cut, and (III) photometric classification cut to the variable objects detected in the background region (1.44 deg$^2$; see Figure \ref{fig:pointing}).  The resulting background rate are summarized in Table \ref{tab:bkg},  where the errors  are evaluated using the Poisson statistics. These are fully evaluated with the data-driven method.
We stress that we do not perform any simulated observations with lightcurve templates at this stage. 

\begin{deluxetable*}{lcc}
\tablewidth{0pt}
\tablecaption{Summary of our background rate estimate by using the data driven method analyzing the background region. The errors are 1-$\sigma$ Poisson noise. The threshold parameters for photometric classifications are given in Table \ref{tab:param}. See text for the estimation of the values in the last row.  }
\tablehead{
  Selection &   $N_{\rm bkg}~^a$  & $\tilde{\mu}_{\rm bkg}~^b$  \\
            &  (1.44 deg$^2$)           &         (0.918 deg$^2$)    
}
\startdata
   Total       &   8359  $\pm$ 91    &       5329 $\pm$ 58      \\      
   Image-quality cut       &      567 $\pm$ 24    &      361 $\pm$ 15        \\      
   Galactic star cut  &  293 $\pm$ 17      &    187 $\pm$ 11   \\ 
   Photometric classification cut &  0   & 0  \\     
  Expected photometric classification cut  & (0.048 $\pm$ 0.038) & (0.030 $\pm$ 0.024)
\enddata
\tablecomments{  
  $^a$ Total number of TDE candidates in the background region of 1.44 deg$^2$. \\
  $^b$ The expected number of background events scaled to the covered region of 0.918 deg$^2$  after each selection process. \\
}
\label{tab:bkg}
\end{deluxetable*}

\subsubsection{Image-quality cut}

The number of sources detected after the image subtraction in the background region is 8359 ($\pm$ 91).
After imposing the image-quality cut,
the number of remaining sources is 567 ($\pm 24$).
This corresponds to the background rate of 
361 ($\pm 15$) sources per our source region (0.918 ${\rm deg^{2}}$).

\subsubsection{Exclusion of Galactic Stars}

We apply our Galactic star cuts to the remaining objects that passes the image-quality cut. 262 objects are removed by the negative pixel evaluation, and further 12 objects are matched with the Gaia variable source catalog. These cuts leave 293 ($\pm$ 17) objects in the background region, which corresponds to 187 $\pm$ 11 in the source region.

\subsubsection{Photometric classification}

\begin{deluxetable}{lcc}
\tablewidth{0pt}
\tablecaption{Threshold parameters for (III) photometric classification cut. The same values of $\mathcal{R}_{\rm Mm,th}$ and $\mathcal{R}_{\rm color,th}$ are used in all the bands for simplicity.} 
\tablehead{
 ${\mathcal{R}_{\rm Mm,th}}^a$ & ${\mathcal{R}_{\rm color,th}}^b$ & ${\mathcal{R}_{g,\rm host,th}}^c$
}
\startdata
1.0 & 1.2 & 3.5
\enddata
\tablecomments{  
  $^a$ The flux ratio at the maximum time to the minimum time among the epochs (criterion (5)). \\
  $^b$  The flux ratio of the consecutive bands (criterion (6)).\\
  $^c$ The flux ratio of the variable object to the host (criterion (7))
}
\label{tab:param}
\end{deluxetable}

To apply photometric classification, we need to tune the threshold parameters introduced in Section \ref{sec:selection_overview} such that the ratio of the TPR to the background rate is maximized (see Section \ref{sec:statistics}). The resulting parameters are given in the caption of Table \ref{tab:param}, and details (rejection rate and number of background by each criterion) are given in Appendix \ref{sec:photometric_detail}.

After applying the photometric classification, we found that the number of remaining objects becomes 0. In order to estimate the expected number of TDE candidate left in the background region, we first evaluate the independency of photometric classification criteria (1) - (6) and criterion (7) and find that these two are independent within the range of our investigation (see Appendix \ref{sec:photometric_detail}). Given the independency of these criteria, the expected number of background region can be estimated by 
\begin{equation}
N_{\rm bkg}^{\rm final}\approx N_{\rm bkg}^{\rm GS}\cdot P_{(1)..(6)}\cdot P_{(7)}    ,\label{eq:Nbkg}
\end{equation}
where $N_{\rm bkg}^{\rm final}$ is the number of background objects after all the selection, $N_{\rm bkg}^{\rm GS}$ is the number of background objects after applying (II) Galactic star cut, $P_{(1)...(6)}$ and $P_{(7)}$ are the passing rate of the criteria (1) - (6) and criterion (7), respectively. This estimate is shown in the last column of Table \ref{tab:bkg}.

In summary, the background objects expected in the source region is estimated to be $0.030\pm0.024$. The large error is caused by a low number of objects that pass criteria (1) - (6) and criterion (7) in (III) photometric classification cut.

\subsection{Estimates of true positive rates}
\label{sec:tpr}
\begin{figure}[t!]
\epsscale{1.0}
\plotone{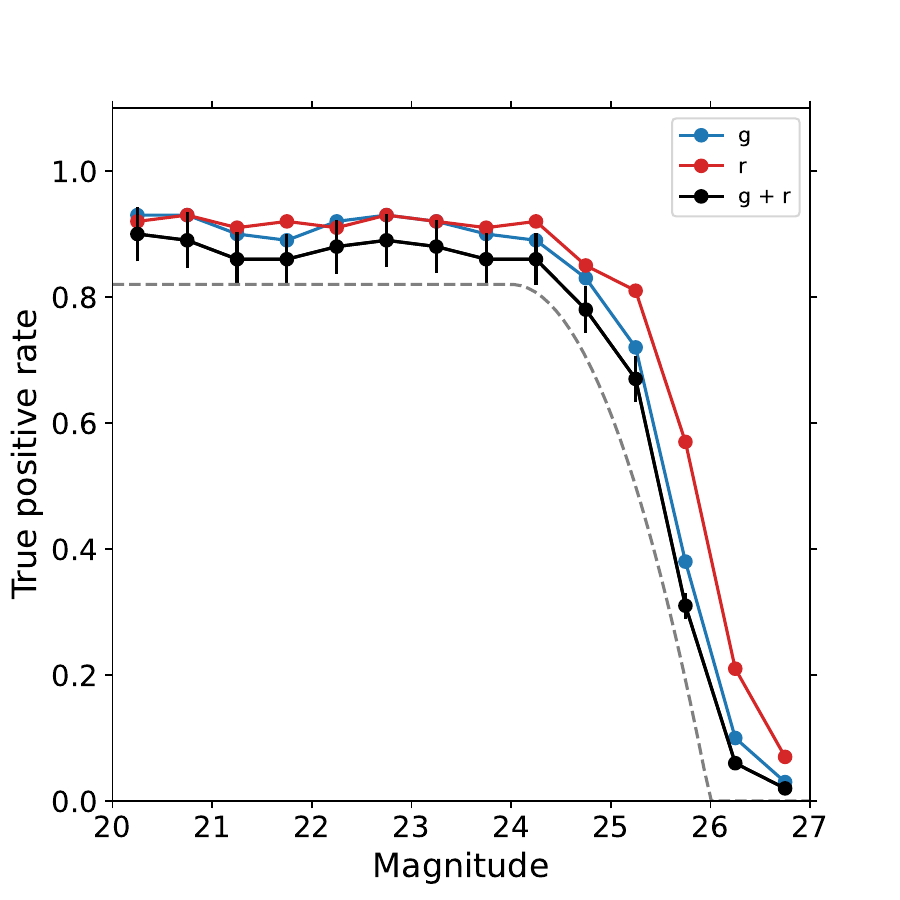}
\caption{TPR of (I) image-quality cut as a function of magnitude. The resulting TPR (black) is measured by imposing two detection in $g$ and $r$ bands at Epoch 1 (with 1 $\sigma$ Poisson noise).
Blue and red points show the TPRs for $g$- and $r$-band detection, respectively. The gray dashed curve shows a smooth function used in the simulated observations in Section \ref{sec:tpr:photo}.} 
\label{fig:tpr}
\end{figure}


\begin{deluxetable}{lccc}
\tablewidth{0pt}
\tablecaption{Summary of our TPR estimate by analysis in the background region and the simulated observations. The 1-$\sigma$ error of our TPR estimate is evaluated by binomial distribution. We estimate it to be $\hat{\sigma}_{\rm TPR}=0.0044$ for the entire neutrino error region, which is scaled to $\sigma_{\rm TPR}=0.0033$ in the covered region. The threshold parameters for photometric classifications are tabulated in Table \ref{tab:param}}
\tablehead{
  Selection & TPR$^a$ & Cum. TPR$^b$  & $\tilde{\mu}_{\rm sig}^c$\\
}
\startdata
   Galactic star cut    &  0.99   &   0.99      &   0.73       \\
   Detection  & 0.97 & 0.96 & 0.70 \\
   Image-quality cut & 0.73   &   0.71    &   0.52    \\  
    Photometric classification cut &  0.21 & 0.15 & 0.11 \\     
\enddata
\tablecomments{  
  $^a$ TPR for individual step \\
  $^b$ Cumulative TPR \\
  $^c$ Expected probability of the source TDE discovery, scaled to the covered probability of 73.5\% after each selection process.
}
\label{tab:tpr}
\end{deluxetable}

Next, we evaluate the TRP for our selection procedure. We construct the TPR for (I) image-quality cut as a function of magnitude with the data-driven method by using the actual HSC data. The TPR for (II) Galactic star cut is also evaluated with the data-driven method. On the other hand, the TPR for (III) photometric classification cut is estimated using the simulation-driven method because we do not have a sufficient number of TDE candidates in our background region. Since the TPR for (I) depends on the observed flux distribution, we implement the TPR for (I) in our simulated observations to evaluate the total TPR of our selection criteria.

\subsubsection{Image-quality cut}
\label{sec:bg:image}

We measure the TPR of image-quality cut by injecting artificial point sources (with 20--27 mag) to the original CCD images of the HSC.
To mimic TDEs, we inject artificial sources at the positions where real sources are detected in the original images.
Then, imaging data with artificial sources are processed in the same manner with the original data, i.e., standard image reduction, image subtraction, and source detection in the subtracted images.
Then, the same image-quality cut is applied to the artificial sources, which gives a TPR of this selection (as a function of magnitude).

Figure \ref{fig:tpr} shows the TPR of (I) image-quality cut as a function of magnitude.
The TPR is about 0.85 at 20--24 mag.
Many artificial sources that are not detected are located near very bright stars, around which the quality of image subtraction is poor.
Near the detection threshold ($\sim 25-26$ mag),
the TPR drops below 0.5.
For the simulated observations performed in Section \ref{sec:tpr:photo},
we use the smoothed curve which matches the lower side of the measured TPR (the gray dashed line in Figure \ref{fig:tpr}).

\subsubsection{Galactic star cut}

Galactic star cut consists of two criteria: one is exclusion of the candidates associated with negative pixels
and the other is the crossmatch with the Gaia variable source catalog.
For the former, we measure the TPR by using the image with artificial point sources (see Section \ref{sec:bg:image}).
The injected sources on top of the existing sources can sometimes cause a large number negative pixels due to imperfect image subtraction.
We find that 1\% of the injected sources are mistakenly excluded by our criterion (more than 10 significant negative pixels in 20 $\times$ 20 pixel region). Thus, the TPR of this criterion is 0.99.

For the crossmatch with the Gaia variable source catalog, 
there is a small probability that the true neutrino source is located on the Gaia variable stars by chance. 
We estimate the chance probability by using a typical number density of Gaia variable stars at a Galactic latitude similar to that of IC230724A ($b=-58.5$ deg). 
The number density in this Galactic latitude is found to be about 23 variable objects/deg$^2$, i.e., 21.1 objects in our source region (0.918 deg$^2$).
Thus, the chance coincidence probability within 2 arcsec radius is $3 \times 10^{-5}$. 
The corresponding TPR of this selection criterion is 0.99997, which is equivalent to 1 for our purpose.  

\subsubsection{Photometric classification}
\label{sec:tpr:photo}

We perform simulated observations of source TDEs using SNCosmo package\footnote{\href{https://sncosmo.readthedocs.io/en/stable/index.html}{SNCosmo: https://sncosmo.readthedocs.io}} \citep{2016ascl.soft11017B}. We construct our own TDE lightcurve templates described in Section \ref{sec:source} and Appendix \ref{sec:TDE_lc}. 
By generating the TDEs\footnote{We do not generate other transients, such as supernovae in the simulated observations, because they are not needed to estimate TPRs.}, we create 6441 TDE lightcurves using the same observation conditions (epoch, filter, sky noise level) as our Subaru/HSC observations. These lightcurves contain discretized values of flux at each photometric band at the time of observation epochs. 
Then, applying (I) image-quality cut and (III) photometric classification, we evaluate the TPR for our selection procedure. 
The TPR for (I) image quality cut depends on the magnitude distribution of neutrino emitting TDEs, and thus, we need to evaluate the TPR with simulated observations.
Also, since the detected objects have different magnitude in $g$ and $r$ bands, we use the TPR of the fainter band when applying (I) image quality cut. 
The results are shown in Table \ref{tab:tpr}, where we see the resulting TPRs for (I) image-quality  and (III) photometric classification cuts are 0.73 and 0.21, respectively.
We estimate the error of TPR using binomial statistics:
\begin{equation}
    \sigma_{\rm TPR}=\frac{\sqrt{N_{\rm left}(1-N_{\rm left}/N_{\rm tot})}}{N_{\rm tot}},
\end{equation}
where $N_{\rm tot}=6441$ is the total number of generated TDEs and $N_{\rm left}=962$ is the number of TDEs left after our selection. With these numbers, we obtain $\sigma_{\rm TPR} = 0.0044$. This number needs to be scaled to our source region value by multiplying the covered fraction of 0.735, leading to  the TPR of our entire selection process to be 0.11 ± 0.0033.

\section{Statistical tests}\label{sec:statistics}

We test the hypothesis that IC230724A originated from a TDE using the likelihood analysis. 
After applying all the cuts discussed in Section \ref{sec:selection}, $\mu_{\rm sig}$ of the signal event and $\mu_{\rm bkg}$ of unrelated transients are left in the field. 
In our setup, the resulting number of the signal transient must be 0 or 1, meaning that we can use binomial probability distribution. The number of the background transient could be some integer, which is determined by Poisson distribution.
In addition, we assume that our source model, a population of TDEs, contributes a fraction of $\lambda$ to the diffuse neutrino background. Then, we can write the likelihood for signal hypothesis as \citep[see][]{2025ApJ...993...23T}
\begin{equation}
\mathcal{L}_{\rm sig}=
\begin{cases}
(1-\lambda\mu_{\rm sig})e^{-\mu_{\rm bkg}} & (n_T = 0) \\
1-(1-\lambda\mu_{\rm sig})e^{-\mu_{\rm bkg}} & (n_T \ge 1)
\end{cases},
\end{equation}
where $n_T$ is the number of transients left after the entire selection procedure.
This likelihood is compared to the one for alternative hypothesis, $\mathcal{L}_{\rm alt}(\hat{\lambda})$, in which $\hat{\lambda}$ is chosen such that the likelihood is maximized:

\begin{equation}
\mathcal{L}_{\rm alt}=
\begin{cases}
e^{-\mu_{\rm bkg}} & (n_T = 0) \\
1-(1-\mu_{\rm sig})e^{-\mu_{\rm bkg}} & (n_T \ge 1)
\end{cases},
\end{equation}
where we use $\hat{\lambda}=0$ and 1 for the case with $n_T=0$ and 1, respectively. Then, the log-likelihood ratio, or our test statistic, is given by 
\begin{equation}
    \Lambda=-2\ln\left(\frac{\mathcal{L}_{\rm sig}}{\mathcal{L}_{\rm alt}}\right)
\end{equation}

We evaluate the probability of realizing the cases with $n_T=0$ or $n_T=1$ using the distribution of test statistic, which is evaluated by performing $10^5$ mock observations based on our signal hypothesis by taking into account the estimated errors.
We note that these mock observations, which is a distinct procedure from the simulated observations in Section \ref{sec:tpr:photo}, can be performed if we provide the values of $\mu_{\rm sig}$ and $\mu_{\rm bkg}$, and we do not need to lightcurve information at this stage.
The allowed regions of $\lambda$ is evaluated by the method of \citet{1998PhRvD..57.3873F}.

The threshold parameters for (III) photometric classification cut (see Section \ref{sec:selection}) is chosen such that $\Lambda$ is maximized with the conditions of $n_T=1$ and $\lambda=0$. In such a case, we can write 
\begin{equation}
\frac{\mathcal{L}_{\rm sig}}{\mathcal{L}_{\rm alt}} = \frac{1-\exp(-\mu_{\rm bkg})}{1-(1-\mu_{\rm sig})\exp(-\mu_{\rm bkg})} \approx \frac{\mu_{\rm bkg}}{\mu_{\rm sig}+\mu_{\rm bkg}},   
\end{equation}
where we use $\mu_{\rm bkg}\ll1$ for the last equation. Then, maximizing $\Lambda$ is equivalent to maximizing $\mu_{\rm sig}/\mu_{\rm bkg}$. Thus, our parameter choice is equivalent to maximizing the purity of the signal.
We think that this choice is maximizing the chance to identify an association. This explanation is counter-intuitive because this choice leads to a low value of TPR $\simeq0.11)$. Nevertheless, our results have a lower value of $\mu_{\rm bkg}\simeq 0.03$, and we may be able to put constraint on $\lambda$ with a $\sim90\%$ confidence level if we found a candidate. If we used weaker photometric classification cut criteria, we would have a higher chance of finding a candidate. However, we could not judge whether this candidate is signal or background because both $\mu_{\rm sig}$ and $\mu_{\rm bkg}$ would be high. Such a situation needs to be avoided by optimizing the photometric classification criteria.

\section{Results for follow-up observations to IC230724A}
\label{sec:result}

\begin{figure}[t!]
\epsscale{1.0}
\plotone{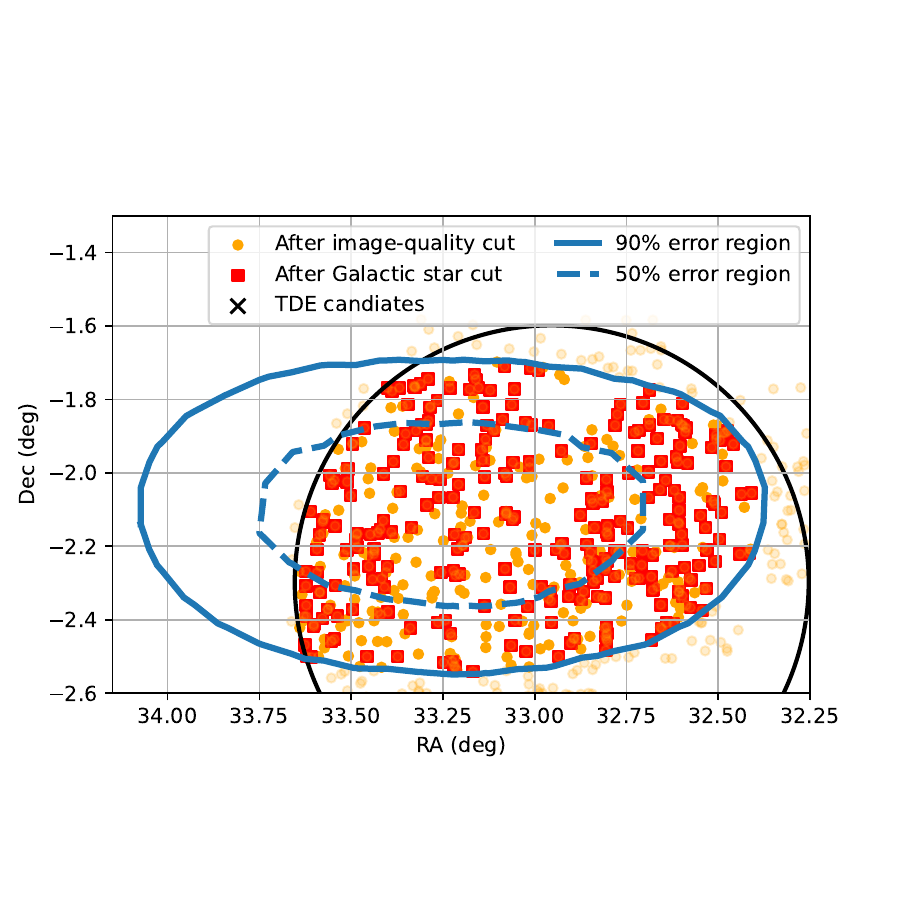}
\caption{Variable objects in our source region. The solid- and dashed-blue lines are the 90\% and 50\% uncertainty ranges of IC230724A, respectively. Our source region is the region enclosed by both the black- and blue-solid lines. The orange circles and red squares are locations of variable objects left after (I) image-quality cut and (II) Galactic star cut, respectively. No object is left after (III) photometric classification cut.
} 
\label{fig:unblind}
\end{figure}

In Section \ref{sec:selection}, we have determined the values of TPR, $\mu_{\rm sig}$, and the background rate, $\mu_{\rm bkg}$, by thorough analysis in the background region. Then, using the statistical method given in Section \ref{sec:statistics}, we have analyzed the data in our source region. As a result, no TDE candidate is found in our source region after the final photometric classification cut. The results are summarized in Table \ref{tab:result} and Figures \ref{fig:unblind}, where orange circles and red squares are variable objects that pass (I) image-quality cut and (II) Galactic star cut, respectively. After (III) photometric classification cut, no objects are left in the source region\footnote{We notice that 
the numbers of transient candidates
given in Table \ref{tab:result} are higher than the numbers estimated in the background region given in Table \ref{tab:bkg}. This might be caused by vignetting, which causes slight degrading in the limiting magnitude ($\sim 0.1$ mag) at the edge of the field-of-view, leading to $\sim$30\% decrease in the number of unrelated transients. Our analysis result is not affected by this slightly increased background rate as the Poisson uncertainty in the background rate is more dominant.}.

\begin{figure}[t!]
\epsscale{1.0}
\plotone{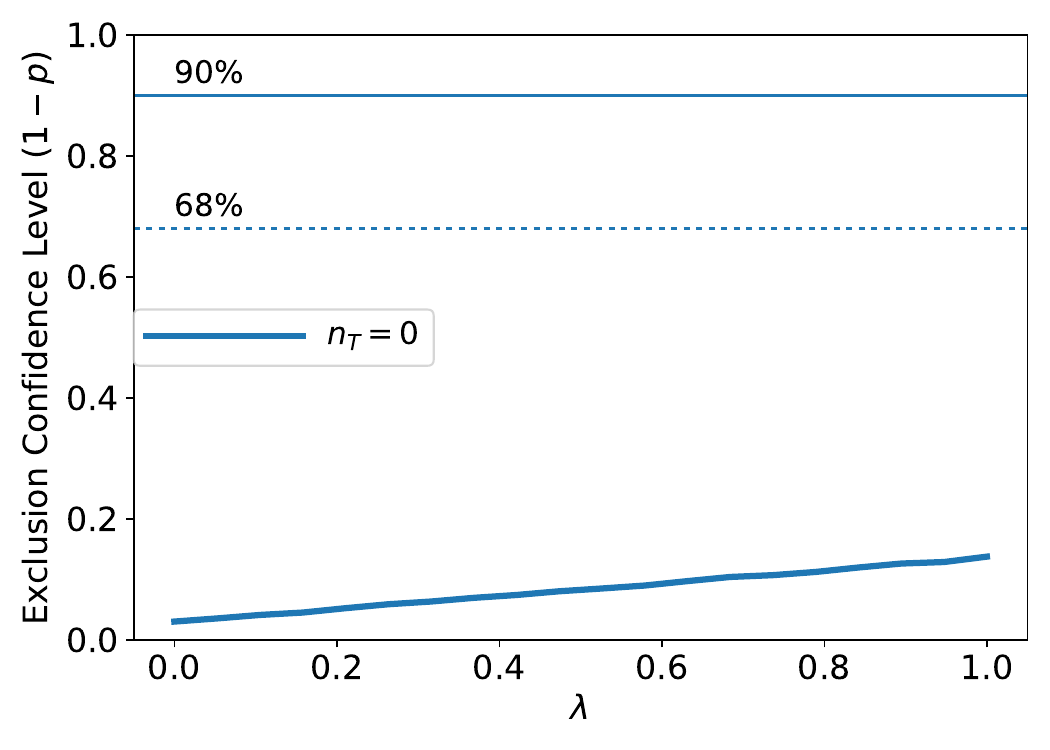}
\caption{The excluded confidence level, $1-p$, for $n_T=0$ (thick-blue-solid line) as a function of $\lambda$.  Since the probability of realizing $n_T=0$ is always higher than 0.1, all the range, $0\le\lambda\le1$, is allowed.} 
\label{fig:prob_lambda}
\end{figure}

We would like to discuss constraint on the fractional contribution of TDEs to the cosmic neutrino background, $\lambda$. Since our analysis results in no TDE candidate, we perform $10^5$ mock observations for a given value of $\lambda$. Using these and statistical method discussed in Section \ref{sec:statistics}, we compute the probability of realizing $n_T=0$ as a function of $\lambda$. If the probability is less than 0.1, $\lambda$ could be constrained with the 90\% confidence level. 
In other words, we can write the exclusion confidence level by 1-probability. 
Our result shown in Figure \ref{fig:prob_lambda} indicates that the probability is always higher than 32\%, and thus, we do not put any constraint on $\lambda$ by the current dataset even with a 68\% confidence level.

\begin{deluxetable}{lcc}
\tablewidth{0pt}
\tablecaption{Number of variable objects within our source region.}
\tablehead{
  Selection &   $\tilde{n}_T~^a$    \\
}
\startdata
   Total       &   5800      \\      
   Image-quality cut       &    485       \\      
   Galactic star cut   &  275     \\ 
   Photometric classification cut &  0    \\     
\enddata
\tablecomments{  
  $^a$ Total number of TDE candidates in our source region of 0.918 deg$^2$  after each selection process. \\
}
\label{tab:result}
\end{deluxetable}

\section{Future prospects with Rubin/LSST}
\label{sec:rubin}

Although our follow-up observations to a single neutrino event cannot constrain the source TDE population, multiple follow-up campaigns near future will be able to put a meaningful constraint. In this section, we discuss the prospects for constraints on the source TDE population with Rubin/LSST \citep{2019ApJ...873..111I}, which will provide an excellent photometric data set. These data sets should cover the entire error regions of multiple neutrino alerts. Here, we evaluate the expected constraint on the fractional contribution of TDEs to the cosmic neutrino background.

In case we perform our statistical analysis on $N_{\rm trial}$ neutrino alerts, the likelihood for signal and alternative hypotheses can be modified to
\begin{align}
\mathcal{L}_{N,\rm sig}(N_{n_T\ge1};~\lambda)=\prod_i^{N_{\rm trial}} \mathcal{L}_{{\rm sig},i}\\
\mathcal{L}_{N,\rm alt}(N_{n_T\ge1};~\hat{\lambda})=\prod_i^{N_{\rm trial}} \mathcal{L}_{{\rm alt},i}
\end{align}
where subscript $i$ indicates the quantities on the $i$th trial and $N_{n_T\ge1}$ is the number of trials 
in which more than 1 objects are found, i.e., the number of trials that satisfy $n_{T,i}\ge1$. 
We should note that $\mathcal{L}_{N,\rm sig}$, $\mathcal{L}_{N,\rm alt}$, and $\hat{\lambda}$ depend only on $N_{n_T\ge1}$. With these likelihood, we can write the test statistic as
\begin{equation}
\Lambda_N(N_{n_T\ge1})=-2\ln\left(\frac{\mathcal{L}_{N,\rm sig}}{\mathcal{L}_{N,\rm alt}}\right)
\end{equation}

\begin{figure}[t!]
\epsscale{1.0}
\plotone{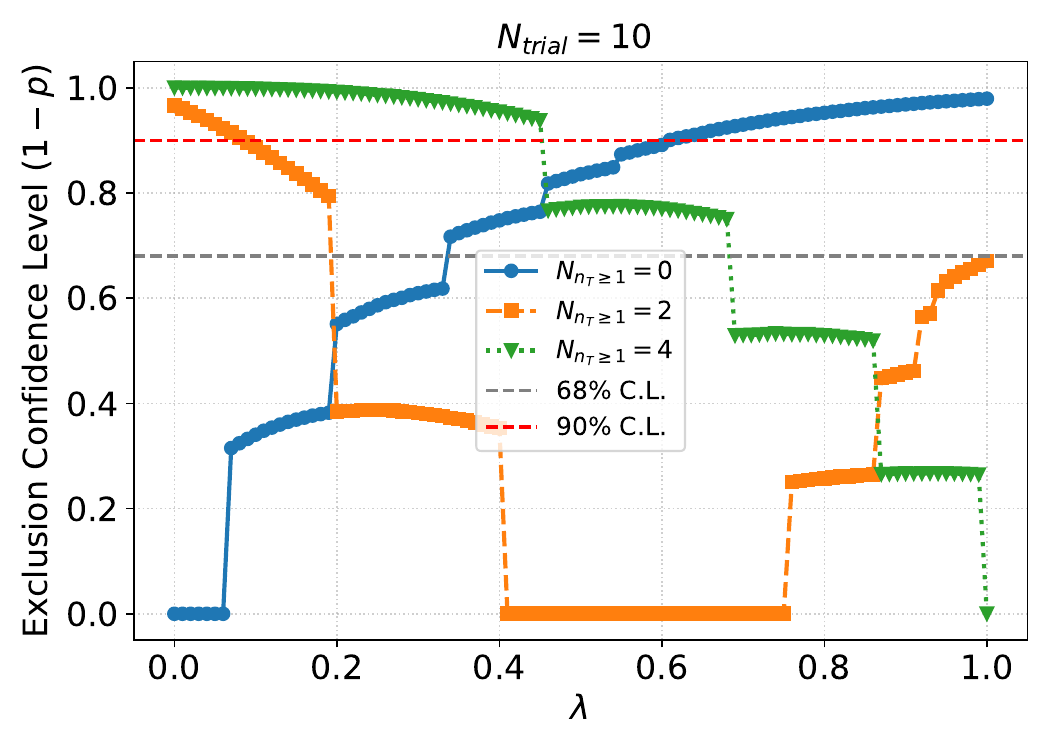}
\plotone{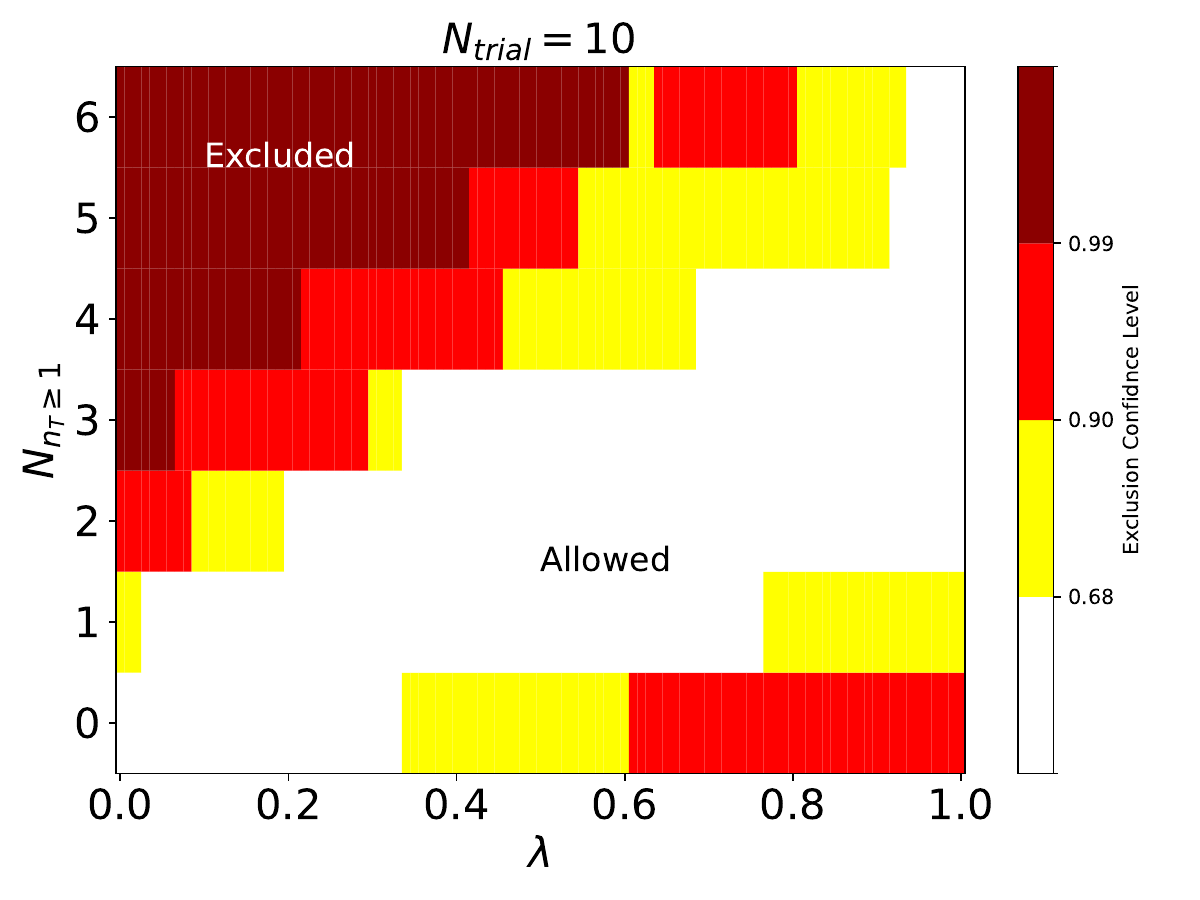}
\plotone{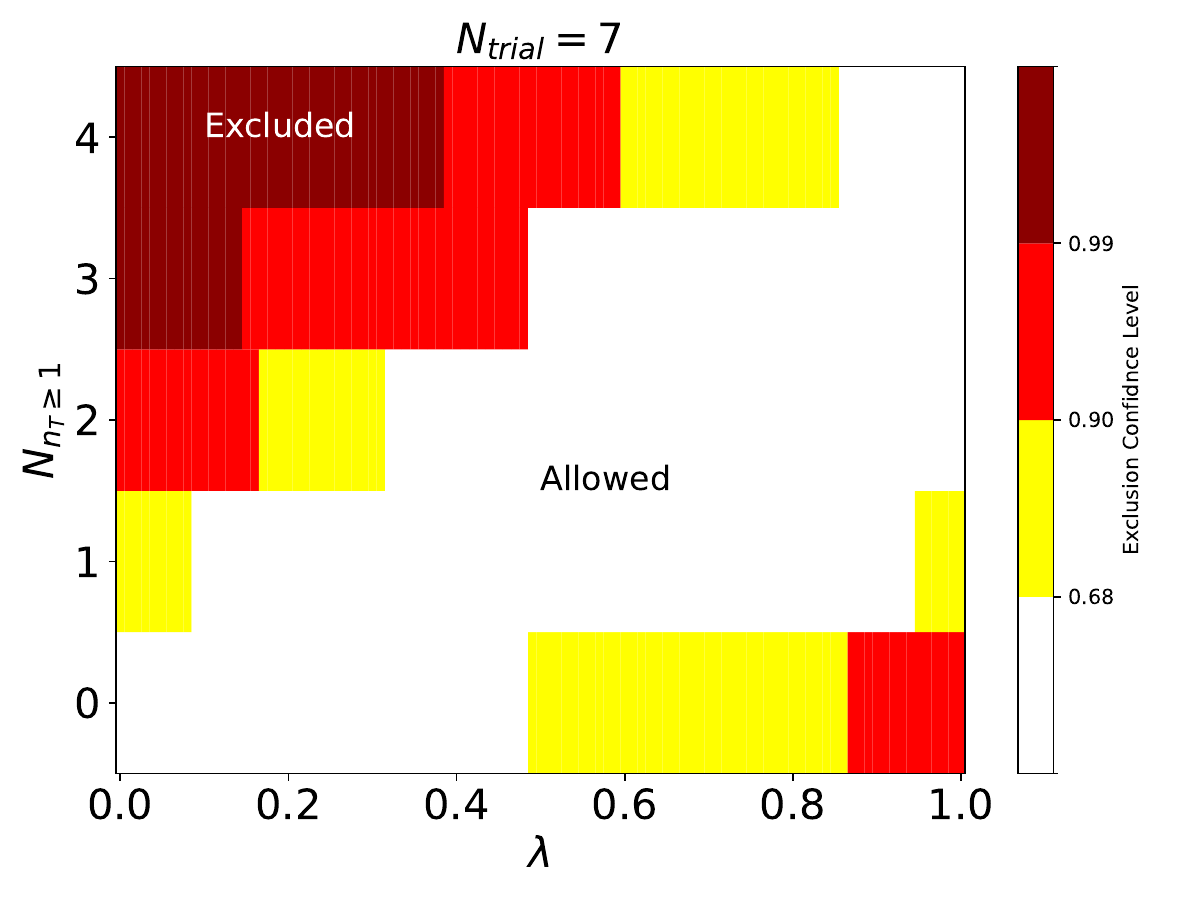}
\caption{Expected constraints on the fractional contribution of TDEs, $\lambda$, with Rubin/LSST. The top panel shows the exclusion confidence level on $\lambda$ for various values of $N_{n_T\ge1}$ for $N_{\rm trial}=10$. The middle and bottom panels show the excluded regions on the $\lambda-N_{n_T\ge1}$ plane for $N_{\rm trial}= 10$ and 7, respectively.}
\label{fig:rubin}
\end{figure}

Since Rubin/LSST will have 10-year base-line data, we can safely exclude steady variable objects, including Galactic variable stars, blazars, and AGNs. Thus, we can focus on photometric classification of TDEs and SNe, which is much easier than what we did in Section \ref{sec:selection}. As reference values, we conservatively assume TPR=0.5 and $\mu_{\rm bkg} = 0.03$, although recent photometric TDE classifiers might be able to achieve better scores \citep{2025A&A...703A..95B,2024ApJ...965L..14S}. Considering signalness = 0.6 for high-quality IceCube Gold alerts, we set $\mu_{\rm sig}=0.3$ for our analysis. IceCube issues Gold Alerts a few times a year, and roughly half of them would be observable by Rubin/LSST for a few months\footnote{The IceCube alerts are mostly concentrated on the equatorial plane of the sky at $|\delta|\lesssim15^\circ$ where Rubin/LSST will be able to observe regularly.}. Considering the 10-year duration of Rubin/LSST, we expect $N_{\rm trial}=5-10$. With these numbers of $\mu_{\rm sig}$, $\mu_{\rm bkg}$, and $N_{\rm trial}$, we estimate the constraints on the fractional contribution of TDE to the diffuse neutrino background, $\lambda$.

The results are shown in Fig. \ref{fig:rubin}. The top panel indicates the exculsion confidence level, $1-p$, as a function of $\lambda$ for $N_{n_T\ge1}=0$, 2, 4. We can see that at least some values of $\lambda$ can be constrained with these numbers of $N_{n_T\ge1}$. The middle and bottom panels indicate the excluded regions on the $\lambda-N_{n_T\ge1}$ plane.
With 10 trials ($N_{\rm Trial}=10$; bottom panel), we will be able to exclude $\lambda\lesssim0.3$ if we detect TDE candidates three times or more ($N_{n_T\ge1}\ge3$). On the other hand, $\lambda\gtrsim 0.6$ would be excluded if we do not detect any TDE candidate ($N_{n_T\ge1}=0$), i.e., the fractional contribution of the TDE population should be lower than 60\%.
With 7 trials ($N_{\rm Trial}=7$; top panel), we will be able to exclude $\lambda\lesssim 0.2$ with 90\% confidence level if we detect TDE candidates twice or more ($N_{n_T\ge1}\ge2$). This is equivalent to state that the fractional contribution by the TDE population should be larger than 20\%. In contrast, $\lambda\gtrsim0.85$ would be excluded if we do not detect any TDE candidate ($N_{n_T\ge1}=0$). 

We stress that this constraint is achievable without dedicated ToO observations by Rubin\footnote{Spectroscopic ToO observations with other telescopes are helpful to firmly classify transients as discussed in Section \ref{sec:summary}.}: Our estimates adopt the values of $\mu_{\rm sig}$ and $\mu_{\rm bkg}$ for the main survey mode (Wide-Fast-Deep; WFD) of LSST \citep{2022ApJS..258....1B}\footnote{See \href{https://zenodo.org/records/15128504}{https://survey-strategy.lsst.io} for the latest survey strategy}. 

Our estimates are conservative in the sense that we do not consider neutrino events from near-future experiments, including KM3NeT \citep{2024EPJC...84..885K}. These experiments will increase the number of well-localized neutrino events in the Southern Hemisphere, which is monitored by Rubin. If higher values of $N_{\rm trial}$ is achieved, we will be able to get tighter constraints.

\section{Summary \& Discussion}
\label{sec:summary}

We performed deep Subaru/HSC follow-up observations to IC230742A and examined a TDE hypothesis as the source of the neutrino event. We adopt the blind analysis policy to exclude bias, which enables us to robustly constrain or confirm the TDE hypothesis. We found that our current follow-up data are insufficient to constrain the hypothesis. This is partly because we had the limited number of observational epochs that prevents us from isolating the TDEs from other variables/transients. If we had 4-5 observational epochs for total duration of a few months, we would be able to distinguish TDEs from other variables as good as $\rm TPR\sim0.5$ and $\mu_{\rm bkg}\lesssim0.1$, based on our simulated observations. 
Nevertheless, we could not conclusively constrain or confirm the TDE hypothesis by follow-up observations to a single neutrino event even if ${\rm TPR\sim0.5}$ was satisfied. Follow-up observations to multiple neutrino events are essential to firmly identify or exclude the TDE population as neutrino sources.

We also examined future prospects with Rubin/LSST and found that we will be able to constrain or confirm the TDE hypothesis with the data taken by its main survey mode. 
The fractional contribution to cosmic neutrino background will be constrained either $\lambda\gtrsim0.6$ or $\lambda\lesssim0.3$ for non-detection or detection, respectively, if 10 neutrino error regions are covered by Rubin/LSST survey.
Our results have significant implications for the follow-up strategy in the Rubin era; Robust constraints by the blind analysis can be obtained without performing ToO observations. 

Rubin will perform ToO observations to excellent neutrino events. ToO observations will provide higher cadence and deeper data, which should increase the TPR of our analysis. Nevertheless, we need to be cautious when estimating the number of background objects. ToO observations will cause non-uniform data quality, which most likely reduces the area of the background region that has cadence and depth similar to the source region. This may cause to increase the uncertainty of the number of background objects, and then, constrain could be weaker than that obtained by usual survey data. 
In order to fully utilize the good-quality ToO data,
it is desirable to also observe a background region with the same data quality (cadence, depth, photometric bands)
over $\sim20-30\rm~deg^2$, which enables accurate background estimation (see Appendix \ref{sec:photometric_detail}).
This can be achieved with a few additional pointings surrounding the source region thanks to a large FoV of Rubin Observatory.

In this study, we ignore TDEs happening in AGN. Although some TDE-candidates are reported in AGN environments \citep[e.g.,][]{2021ApJ...920...56F,2022PhRvL.128v1101R}, evaluation of TDE occurence rates in AGN is still challenging because of variable nature and high optical/UV luminosity in AGN environments. Some theoretical studies suggest that TDEs in AGN environments would potentially produce neutrinos more efficiently, e.g., due to cloud-outflow interaction \citep{2022MNRAS.514.4406W}. Careful investigation of neutrino production in TDEs happening in AGN is left as future work. 

We focus on the TDE population in our analysis because it is rare and easy to identify/exclude as neutrino sources. Some specific types of supernovae, such as broad-line Type Ic SNe, Type IIn SNe, and superluminous supernovae are also proposed as neutrino sources \citep[e.g.][]{MuraseIoka13a,2011PhRvD..84d3003M,2017MNRAS.470.1881P,2014PhRvD..90j3005F,Mukhopadhyay2026arXiv}. Photometric classification of these specific sub-types of supernovae is harder than the TDE-SNe classification. We defer the detailed examinations of the photometric classifications among various SN types and prospects for constraints on these SNe as neutrino sources with Rubin/LSST to future work.

Spectroscopic observations of transients are necessary to solidly classify variable objects, including SNe sub-types. However, it is not easy to perform spectroscopic observations to all the variable objects found in the error region, $\sim50-200$ objects with $\sim22 - 24$ mag spreads in $\sim 1$ deg$^2$. Wide-field multi-object spectroscopic observations are necessary to make this strategy feasible, which is currently achievable only with Subaru/\lq\={O}nohi\lq ula Prime Focus Spectrograph \citep{2016SPIE.9908E..1MT}.

Future water-based neutrino detectors will achieve better angular resolution with $\lesssim0.1$ degree \citep[e.g.,][]{2020NatAs...4..913A,2023NatAs...7.1497Y,2024EPJC...84..885K}. Then, the number of transients within error regions is expected to be less than 10 objects, enabling us to perform follow-up observations with narrower field-of-view telescopes. However, the expected distances to the source for current neutrino alerts are cosmological, $z\sim0.3-1$, requiring 8-m or larger telescopes for spectroscopic observations. The distance to the source should be constrained by a nearby Universe with the help of the newly developed multiplet alert system \citep{2025ApJ...981..159A}. This  enables us to focus on bright transients with widely available 1-2 m telescopes \citep{2025ApJ...993...23T}. Combining these updated neutrino alert systems and  strategic optical surveys, we will be able to identify or rule out optical transients as neutrino sources near future.

\begin{acknowledgments}
We thank Shigeru Yoshida and Aya Ishihara for fruitful discussion.

This work is supported by Grant-in-Aid for Scientific research from JSPS for Transformative Research Areas (A) (grant Nos. 23H04891, 23H04892, 23H04894, 23H04899). 
and JSPS KAKENHI Nos. 26K00733, 26K00696 (S.S.K.) 25KJ0556 (S.T.). S.S.K. acknowledges support by  the Tohoku Initiative for Fostering Global Researchers for Interdisciplinary Sciences (TI-FRIS) of MEXT’s Strategic Professional Development Program for Young Researchers. 
S.T. acknowledges support from the Graduate Program on Physics for the Universe (GP-PU) at Tohoku University.

This research is based in part on data collected at the Subaru Telescope, which is operated by the National Astronomical Observatory of Japan. We are honored and grateful for the opportunity of observing the Universe from Maunakea, which has the cultural, historical, and natural significance in Hawaii.
\end{acknowledgments}

%

\vspace{5mm}
\facilities{Subaru HSC \citep{miyazaki06,HSC2018M}}


\software{astropy \citep{2013A&A...558A..33A,2018AJ....156..123A},  SNCosmo \citep{2016ascl.soft11017B}, hscpipe \citep{HSC-pipe2018PASJ}}



\appendix

\section{Glossary of variables}
\label{sec:param_list}

The variables used in our analyses are tabulated in Table \ref{tab:variables}.

\begin{deluxetable*}{r|l}
\tablewidth{0pt}
\tablecaption{ List of variables.}
\label{tab:variables}
\tablehead{    TDE Lightcurve templates
}
\startdata
    $L_{\rm pk}$     & the peak luminosity \\
     $t_{\rm rise}$ & rise timescale \\
     $t_{\rm decay}$ & decay timescale\\
     $t_{\rm pk}$ & peak time of the optical lightcurve \\
     $p_{\rm decay}$ & the decay index\\
     $T_{\rm pk} $    & the peak temperature \\
     $dT/dt$ & time derivative of temperature\\
    \hline
     Photometric classification\\
     \hline
    $\mathcal{R}_{\rm Mm,thr}$ & threshold for flux ratio among epochs for photometric classification criterion (5) \\
     $\mathcal{R}_{\rm color,thr}$ &  threshold for the flux ratio of consecutive bands for photometric classification criterion (6) \\
    $\mathcal{R}_{g,\rm host,thr}$ & threshold for the ratio of variable flux to the host for photometric classification criterion (7)\\
    \hline
     Statistical analysis \\
    \hline
    $\lambda$ & fractional contribution of the TDE population to the cosmic neutrino background\\
    $\mu_{\rm sig}$ & expected number of signal events\\
    $\mu_{\rm bkg}$ & expected number of background events\\
    $n_T$ & number of TDE candidates after selection process\\
    $N_{\rm Trial}$ & number of trials (number of neutrino events)\\
    $N_{n_T\ge1}$ & number of trials in which more than 1 TDE candidates are detected\\
\enddata
\end{deluxetable*}

\section{Construction method for TDE lightcurve templates} \label{sec:TDE_lc}

The ZTF-detected TDE sample \citep{2023ApJ...942....9H} is parametrized with 6 parameters: $L_{\rm pk}$, $t_{\rm rise}$, $t_{\rm decay}$,  $T_{\rm pk}$, $dT/dt$, and $p_{\rm decay}$. We evaluate the correlation between the two of them, and found that $\ln(L_{\rm pk})$ correlates with $T_{\rm pk}$ and $\ln(t_{\rm rise})$. Similarly, $T_{\rm pk}$ correlates with $dT/dt$, and $\ln(t_{\rm rise})$ correlates with $\ln(t_{\rm decay})$. Also, $\ln (t_{\rm decay})$ correlates with $p_{\rm decay}$. We assume that the correlated variables have linear relation with some intrinsic scatter expressed by Gaussian with a standard deviation. The values of these standard deviations are evaluated from the original TDE sample. By using these correlations, we construct a large number of lightcurve templates. 
We set 26 luminosity bins equally divided in the logarithmic space in the range of $10^{43}\rm~erg~s^{-1}-10^{45.5}\rm~erg~s^{-1}$. We allocate 100 TDE lightcurve templates in each bin, resulting in 2600 TDE lightcurve templates in total. 
We restrict all the parameters within the 2-sigma range because the choices of $>$ 2-$\sigma$ parameters easily lead to unphysical light curve behaviors. 
For example, the median and standard deviation of $\ln(t_{\rm decay})$ are 1.66 and 0.30, respectively, which leads to the range of $t_{\rm decay}\approx12-187$ days.
We also limit $T_{\rm floor} < T(t) < T_{\rm cap}$, where $T_{\rm floor} = 10^4$ K and $T_{\rm cap} = 10^5$ K, in order to match the optical features consistent with the optical observed data (no TDEs  exhibiting $T < T_{\rm floor}$ was found so far, and we cannot detect optical TDEs with $T \gg 10^5$ K). In addition, we add another constraint, $p_{\rm decay} \ge 1.0$ because $p_{\rm decay} < 1.0$ means that the released energy increases with time, which is at odds with the observational data and theoretical prediction. Our TDE lightcurve templates have correlation coefficient matrices that are very similar to the ZTF-detected sample. Box’s M test gives a P-value of 0.653, indicating that the parameter distributions of these two populations can be considered identical. 

\section{Details of photometric classification}
\label{sec:photometric_detail}

We show the background estimates and TPR for our photometric classification in each step in Tables \ref{tab:photometric_bkg} and \ref{tab:photometric_tpr}, respectively. 
For the background rates, the values start after (I) image-quality cut and (II) Galactic star cut. 
For the TPR, we first take into account the Galactic star cut, leading to 0.99 as the initial TPR. Then, we perform simulated observations to evaluate the detection rate by the depth with our HSC data. We then apply (I) image-quality cut and (III) photometric classification to the sample of detected TDEs to obtain the values of the TPR.
We also show the number of candidates in our source region in Table \ref{tab:photometric_source} for completeness.

To estimate the final number of background objects, we evaluate the independency between criteria (1) - (6) and criterion (7). We apply the criteria (1) - (6) to $N_{\rm bkg}^{\rm GS}=293$ objects that passes (I) and (II), and 7 objects are left as shown in Table \ref{tab:photometric_bkg}. This means that the passing rate of criteria (1)-(6) is $P_{(1)...(6)}=0.024$. Independently, we also apply the criterion (7) to $N_{\rm bkg}^{\rm GS}$ objects, and 2 objects are left, meaning that the passing rate by criterion (7) is $P_{(7)}=0.0068$. 
If the two are independent, the number of objects that pass criteria (1)-(6) and fail criterion (7) is estimated to be $N_{(1)...(6)}^{(7)}|_{\rm estimate}=N_{\rm bkg}^{\rm GS}P_{(1)...(6)}(1-P_{(7)})\simeq6.95$. Similarly, the number of objects that pass criterion (7) and fail criteria (1)-(6) is given by  $N^{(1)...(6)}_{(7)}|_{\rm estimate}=N_{\rm bkg}^{\rm GS}(1-P_{(1)...(6)})P_{(7)}\simeq1.95$. These estimates are consistent with our data set: $N_{(1)...(6)}^{(7)}|_{\rm data}=7$ and $N^{(1)...(6)}_{(7)}|_{\rm data}=2$, which means that these two are independent. We confirm the independency of the two by checking the values of $N_{(1)...(6)}^{(7)}$ and $N^{(1)...(6)}_{(7)}$ for various sets of photometric classification threshold parameters.

Lastly, we would like to roughly estimate the background region required to achieve the background rate of $\mu_{\rm bkg}\sim0.001$. Here, we consider two independent cut criteria, similar to Eq. (\ref{eq:Nbkg}), to estimate the background rate. Then, we can write the background rate as 
\begin{equation}
    \mu_{\rm bkg}\approx \frac{N_{\rm bkg}^{\rm GS}P_1P_2}{A_{\rm bkg}} A_{\rm src},
\end{equation}
where $A_{\rm src}$ and $A_{\rm bkg}$ are the areas of the source and background regions, respectively, and $P_1$ and $P_2$ are the probabilities that pass each cut criterion, respectively. $P_1$ and $P_2$ should be much larger than $1/N_{\rm bkg}^{\rm GS}$ to reduce the Poisson errors. Then, we can re-write the equation in the following form:
\begin{equation}
    \frac{A_{\rm bkg}}{A_{\rm src}} \approx \frac{N_{\rm bkg}^{\rm GS}P_1P_2}{\mu_{\rm bkg}}\gg \frac{1}{\mu_{\rm bkg}N_{\rm bkg}^{\rm GS}}.
\end{equation}
From our analysis, we found that the number of background objects can be roughly written as $N_{\rm bkg}^{\rm GS}\sim300 (A_{\rm bkg}/\rm deg^2)$.  To achieve $\mu_{\rm bkg}<0.001$ with $A_{\rm src}\sim1\rm~deg^2$, $A_{\rm bkg}\gg 2 \rm~deg^2$ should be satisfied. Thus, $A_{\rm bkg}/A_{\rm src}\sim20-30{\rm~deg^2}$ is desirable to significantly reduce and/or accurately estimate the background rate. We should note that this number is not universal; The required area of the background region depends on the depth of the data and area of the source region.

\begin{deluxetable*}{lccc}
\tablewidth{0pt}
\tablecaption{Number of objects during (III) photometric classification cut. We start from the number of objects after (I) image-quality cut and (II) Galactic star cut.}
\label{tab:photometric_bkg}
\tablehead{Selection Criteria & $N_{\rm bkg}$ & $\tilde{\mu}_{\rm bkg}$ & Rejection rate\\
& (1.44 deg$^2$) & (0.918 deg$^2$)
}
\startdata
Objects after (I) and (II)  & 293 & 187 & N/A\\
(1) 5-$\sigma$ detection at $t_1$  & 185 & 118 & 0.369 \\
(2) Long duration & 174 & 111 & 0.059\\   
(3) Monotonically declining  & 76 & 48 & 0.563 \\
(4) Non V-shape SED  & 68 & 43 & 0.105\\
(5) Significant variation among epochs  & 68 & 43 & 0.00\\
(6) Blue color at all epochs & 7& 4.5 & 0.897\\
(7) High flux ratio to host galaxies & 0& 0 & 1.00\\
\enddata
\end{deluxetable*}

\begin{deluxetable*}{lcc}
\tablewidth{0pt}
\tablecaption{TPR obtained by our simulated observations of (I) image-quality cut and (III) photometric classification cut.}
\label{tab:photometric_tpr}
\tablehead{
  Selection Criteria  & TPR for each level  & Cumulative TPR    \\
}
\startdata
Objects after (II)  & 0.99 & 0.99 \\
Detection fraction & 0.972 & 0.962\\
(I) Image-quality cut & 0.733 & 0.706\\
\hline
(III) Photometric classification \\
\hline
(1) 5-$\sigma$ detection at $t_1$  & 1.0 & 0.706 \\
(2) Long duration & 0.999 & 0.705 \\   
(3) Monotonically declining  & 0.990 &0.698 \\
(4) Non V-shape SED  &  0.991& 0.692\\
(5) Significant variation among epochs  & 1.0  & 0.692\\
(6) Blue color at all epochs & 0.423&  0.292\\
(7) High flux ratio to host galaxies & 0.506& 0.148\\
\enddata
\end{deluxetable*}

\begin{deluxetable}{lcc}
\tablewidth{0pt}
\tablecaption{The values of TDE candidates in our source region during photometric classification cut. }
\tablehead{
  Selection &   $\tilde{n}_T~^a$  & Rejection rate  \\
}
\startdata
Objects after (I) and (II) & 275  & N/A \\
(1) 5-$\sigma$ detection at $t_1$  & 173 & 0.371\\
(2) Long duration & 143 & 0.173 \\   
(3) Monotonically declining  & 76 & 0.469 \\
(4) Non V-shape SED  &  67 & 0.118\\
(5) Significant variation among epochs  & 67 & 0.0 \\
(6) Blue color at all epochs & 3 & 0.955\\
(7) High flux ratio to host galaxies & 0 & 1.0\\
\enddata
\tablecomments{  
  $^a$ Total number of TDE candidates in our source region of 0.918 deg$^2$ after each selection process. \\
}
\label{tab:photometric_source}
\end{deluxetable}


\bibliography{sample631}{}

@INCOLLECTION{2023ecnp.book..433K,
       author = {{Kimura}, Shigeo S.},
        title = "{Neutrinos from Gamma-Ray Bursts}",
    booktitle = {The Encyclopedia of Cosmology. Set 2: Frontiers in Cosmology. Volume 2: Neutrino Physics and Astrophysics},
         year = 2023,
       editor = {{Stecker}, Floyd W.},
        pages = {433-482},
          doi = {10.1142/9789811282645_0009},
       adsurl = {https://ui.adsabs.harvard.edu/abs/2023ecnp.book..433K}
}

@INCOLLECTION{2023ecnp.book..483M,
       author = {{Murase}, Kohta and {Stecker}, Floyd W.},
        title = "{High-Energy Neutrinos from Active Galactic Nuclei}",
    booktitle = {The Encyclopedia of Cosmology. Set 2: Frontiers in Cosmology. Volume 2: Neutrino Physics and Astrophysics},
         year = 2023,
       editor = {{Stecker}, Floyd W.},
        pages = {483-540},
          doi = {10.1142/9789811282645_0010},
       adsurl = {https://ui.adsabs.harvard.edu/abs/2023ecnp.book..483M}
}

@INCOLLECTION{2023ecnp.book..107H,
       author = {{Halzen}, Francis and {Kheirandish}, Ali},
        title = "{IceCube and High-Energy Cosmic Neutrinos}",
    booktitle = {The Encyclopedia of Cosmology. Set 2: Frontiers in Cosmology. Volume 2: Neutrino Physics and Astrophysics},
         year = 2023,
       editor = {{Stecker}, Floyd W.},
        pages = {107-235},
          doi = {10.1142/9789811282645_0005},
       adsurl = {https://ui.adsabs.harvard.edu/abs/2023ecnp.book..107H}
}

@ARTICLE{2023ApJ...953L..12J,
       author = {{Jiang}, Ning and {Zhou}, Ziying and {Zhu}, Jiazheng and {Wang}, Yibo and {Wang}, Tinggui},
        title = "{Two Candidate Obscured Tidal Disruption Events Coincident with High-energy Neutrinos}",
      journal = {\apjl},
         year = 2023,
        month = aug,
       volume = {953},
       number = {1},
          eid = {L12},
        pages = {L12},
          doi = {10.3847/2041-8213/acebe3},
archivePrefix = {arXiv},
       eprint = {2307.16667},
 primaryClass = {astro-ph.HE},
       adsurl = {https://ui.adsabs.harvard.edu/abs/2023ApJ...953L..12J}
}

@ARTICLE{2024ApJ...969..136Y,
       author = {{Yuan}, Chengchao and {Winter}, Walter and {Lunardini}, Cecilia},
        title = "{AT2021lwx: Another Neutrino-coincident Tidal Disruption Event with a Strong Dust Echo?}",
      journal = {\apj},
         year = 2024,
        month = jul,
       volume = {969},
       number = {2},
          eid = {136},
        pages = {136},
          doi = {10.3847/1538-4357/ad50a9},
archivePrefix = {arXiv},
       eprint = {2401.09320},
 primaryClass = {astro-ph.HE},
       adsurl = {https://ui.adsabs.harvard.edu/abs/2024ApJ...969..136Y}
}

@ARTICLE{2005ARNPS..55..141K,
       author = {{Klein}, Joshua R. and {Roodman}, Aaron},
        title = "{Blind Analysis in Nuclear and Particle Physics}",
      journal = {Annual Review of Nuclear and Particle Science},
         year = 2005,
        month = dec,
       volume = {55},
       number = {1},
        pages = {141-163},
          doi = {10.1146/annurev.nucl.55.090704.151521},
       adsurl = {https://ui.adsabs.harvard.edu/abs/2005ARNPS..55..141K}
}

@software{2016ascl.soft11017B,
       author = {{Barbary}, Kyle and {Barclay}, Tom and {Biswas}, Rahul and {Craig}, Matt and {Feindt}, Ulrich and {Friesen}, Brian and {Goldstein}, Danny and {Jha}, Saurabh and {Rodney}, Steve and {Sofiatti}, Caroline and et al.},
        title = "{SNCosmo: Python library for supernova cosmology}",
 howpublished = {Astrophysics Source Code Library, record ascl:1611.017},
         year = 2016,
        month = nov,
          eid = {ascl:1611.017},
archivePrefix = {ascl},
       eprint = {1611.017},
       adsurl = {https://ui.adsabs.harvard.edu/abs/2016ascl.soft11017B}
}

@ARTICLE{2022ApJS..258....1B,
       author = {{Bianco}, Federica B. and {Ivezi{\'c}}, {\v{Z}}eljko and {Jones}, R. Lynne and {Graham}, Melissa L. and {Marshall}, Phil and {Saha}, Abhijit and {Strauss}, Michael A. and {Yoachim}, Peter and {Ribeiro}, Tiago and {Anguita}, Timo and {Bauer}, A.~E. and {Bauer}, Franz E. and {Bellm}, Eric C. and {Blum}, Robert D. and {Brandt}, William N. and {Brough}, Sarah and {Catelan}, M{\'a}rcio and {Clarkson}, William I. and {Connolly}, Andrew J. and {Gawiser}, Eric and {Gizis}, John E. and {Hlo{\v{z}}ek}, Ren{\'e}e and {Kaviraj}, Sugata and {Liu}, Charles T. and {Lochner}, Michelle and {Mahabal}, Ashish A. and {Mandelbaum}, Rachel and {McGehee}, Peregrine and {Neilsen}, Jr., Eric H. and {Olsen}, Knut A.~G. and {Peiris}, Hiranya V. and {Rhodes}, Jason and {Richards}, Gordon T. and {Ridgway}, Stephen and {Schwamb}, Megan E. and {Scolnic}, Dan and {Shemmer}, Ohad and {Slater}, Colin T. and {Slosar}, An{\v{z}}e and {Smartt}, Stephen J. and {Strader}, Jay and {Street}, Rachel and {Trilling}, David E. and {Verma}, Aprajita and {Vivas}, A.~K. and {Wechsler}, Risa H. and {Willman}, Beth},
        title = "{Optimization of the Observing Cadence for the Rubin Observatory Legacy Survey of Space and Time: A Pioneering Process of Community-focused Experimental Design}",
      journal = {\apjs},
         year = 2022,
        month = jan,
       volume = {258},
       number = {1},
          eid = {1},
        pages = {1},
          doi = {10.3847/1538-4365/ac3e72},
archivePrefix = {arXiv},
       eprint = {2108.01683},
 primaryClass = {astro-ph.IM},
       adsurl = {https://ui.adsabs.harvard.edu/abs/2022ApJS..258....1B}
}

@ARTICLE{2024MNRAS.534.1528M,
       author = {{Mukhopadhyay}, Mainak and {Bhattacharya}, Mukul and {Murase}, Kohta},
        title = "{Multimessenger signatures of delayed choked jets in tidal disruption events}",
      journal = {\mnras},
         year = 2024,
        month = oct,
       volume = {534},
       number = {2},
        pages = {1528-1540},
          doi = {10.1093/mnras/stae2080},
archivePrefix = {arXiv},
       eprint = {2309.02275},
 primaryClass = {astro-ph.HE},
       adsurl = {https://ui.adsabs.harvard.edu/abs/2024MNRAS.534.1528M}
}

@ARTICLE{2025ApJ...990...18L,
       author = {{Lu}, Ming-Xuan and {Liang}, Yun-Feng and {Wang}, Xiang-Gao and {Ouyang}, Xue-Rui},
        title = "{Investigating the Correlation between ZTF Tidal Disruption Events and IceCube High-energy Neutrinos}",
      journal = {\apj},
         year = 2025,
        month = sep,
       volume = {990},
       number = {1},
          eid = {18},
        pages = {18},
          doi = {10.3847/1538-4357/adef54},
       adsurl = {https://ui.adsabs.harvard.edu/abs/2025ApJ...990...18L}
}

@ARTICLE{HSC-pipe2018PASJ,
       author = {{Bosch}, James and {Armstrong}, Robert and {Bickerton}, Steven and {Furusawa}, Hisanori and {Ikeda}, Hiroyuki and {Koike}, Michitaro and {Lupton}, Robert and {Mineo}, Sogo and {Price}, Paul and {Takata}, Tadafumi and {Tanaka}, Masayuki and {Yasuda}, Naoki and {AlSayyad}, Yusra and {Becker}, Andrew C. and {Coulton}, William and {Coupon}, Jean and {Garmilla}, Jose and {Huang}, Song and {Krughoff}, K. Simon and {Lang}, Dustin and {Leauthaud}, Alexie and {Lim}, Kian-Tat and {Lust}, Nate B. and {MacArthur}, Lauren A. and {Mandelbaum}, Rachel and {Miyatake}, Hironao and {Miyazaki}, Satoshi and {Murata}, Ryoma and {More}, Surhud and {Okura}, Yuki and {Owen}, Russell and {Swinbank}, John D. and {Strauss}, Michael A. and {Yamada}, Yoshihiko and {Yamanoi}, Hitomi},
        title = "{The Hyper Suprime-Cam software pipeline}",
      journal = {\pasj},
         year = 2018,
        month = jan,
       volume = {70},
          eid = {S5},
        pages = {S5},
          doi = {10.1093/pasj/psx080},
archivePrefix = {arXiv},
       eprint = {1705.06766},
 primaryClass = {astro-ph.IM},
       adsurl = {https://ui.adsabs.harvard.edu/abs/2018PASJ...70S...5B}
}

@ARTICLE{HSC2018M,
       author = {{Miyazaki}, Satoshi and {Komiyama}, Yutaka and {Kawanomoto}, Satoshi and {Doi}, Yoshiyuki and {Furusawa}, Hisanori and {Hamana}, Takashi and {Hayashi}, Yusuke and {Ikeda}, Hiroyuki and {Kamata}, Yukiko and {Karoji}, Hiroshi and {Koike}, Michitaro and {Kurakami}, Tomio and {Miyama}, Shoken and {Morokuma}, Tomoki and {Nakata}, Fumiaki and {Namikawa}, Kazuhito and {Nakaya}, Hidehiko and {Nariai}, Kyoji and {Obuchi}, Yoshiyuki and {Oishi}, Yukie and {Okada}, Norio and {Okura}, Yuki and {Tait}, Philip and {Takata}, Tadafumi and {Tanaka}, Yoko and {Tanaka}, Masayuki and {Terai}, Tsuyoshi and {Tomono}, Daigo and {Uraguchi}, Fumihiro and {Usuda}, Tomonori and {Utsumi}, Yousuke and {Yamada}, Yoshihiko and {Yamanoi}, Hitomi and {Aihara}, Hiroaki and {Fujimori}, Hiroki and {Mineo}, Sogo and {Miyatake}, Hironao and {Oguri}, Masamune and {Uchida}, Tomohisa and {Tanaka}, Manobu M. and {Yasuda}, Naoki and {Takada}, Masahiro and {Murayama}, Hitoshi and {Nishizawa}, Atsushi J. and {Sugiyama}, Naoshi and {Chiba}, Masashi and {Futamase}, Toshifumi and {Wang}, Shiang-Yu and {Chen}, Hsin-Yo and {Ho}, Paul T.~P. and {Liaw}, Eric J.~Y. and {Chiu}, Chi-Fang and {Ho}, Cheng-Lin and {Lai}, Tsang-Chih and {Lee}, Yao-Cheng and {Jeng}, Dun-Zen and {Iwamura}, Satoru and {Armstrong}, Robert and {Bickerton}, Steve and {Bosch}, James and {Gunn}, James E. and {Lupton}, Robert H. and {Loomis}, Craig and {Price}, Paul and {Smith}, Steward and {Strauss}, Michael A. and {Turner}, Edwin L. and {Suzuki}, Hisanori and {Miyazaki}, Yasuhito and {Muramatsu}, Masaharu and {Yamamoto}, Koei and {Endo}, Makoto and {Ezaki}, Yutaka and {Ito}, Noboru and {Kawaguchi}, Noboru and {Sofuku}, Satoshi and {Taniike}, Tomoaki and {Akutsu}, Kotaro and {Dojo}, Naoto and {Kasumi}, Kazuyuki and {Matsuda}, Toru and {Imoto}, Kohei and {Miwa}, Yoshinori and {Suzuki}, Masayuki and {Takeshi}, Kunio and {Yokota}, Hideo},
        title = "{Hyper Suprime-Cam: System design and verification of image quality}",
      journal = {\pasj},
         year = 2018,
        month = jan,
       volume = {70},
          eid = {S1},
        pages = {S1},
          doi = {10.1093/pasj/psx063},
       adsurl = {https://ui.adsabs.harvard.edu/abs/2018PASJ...70S...1M}
}

@ARTICLE{2025A&A...703A..95B,
       author = {{Bhardwaj}, Kunal and {Christov}, Asen and {Karpov}, Sergey},
        title = "{A photometric classifier for tidal disruption events in Rubin LSST}",
      journal = {\aap},
         year = 2025,
        month = nov,
       volume = {703},
          eid = {A95},
        pages = {A95},
          doi = {10.1051/0004-6361/202556839},
archivePrefix = {arXiv},
       eprint = {2509.25902},
 primaryClass = {astro-ph.IM},
       adsurl = {https://ui.adsabs.harvard.edu/abs/2025A&A...703A..95B}
}

@ARTICLE{2024ApJ...965L..14S,
       author = {{Stein}, Robert and {Mahabal}, Ashish and {Reusch}, Simeon and {Graham}, Matthew and {Kasliwal}, Mansi M. and {Kowalski}, Marek and {Gezari}, Suvi and {Hammerstein}, Erica and {Nakoneczny}, Szymon J. and {Nicholl}, Matt and et al.},
        title = "{tdescore: An Accurate Photometric Classifier for Tidal Disruption Events}",
      journal = {\apjl},
         year = 2024,
        month = apr,
       volume = {965},
       number = {2},
          eid = {L14},
        pages = {L14},
          doi = {10.3847/2041-8213/ad3337},
archivePrefix = {arXiv},
       eprint = {2312.00139},
 primaryClass = {astro-ph.IM},
       adsurl = {https://ui.adsabs.harvard.edu/abs/2024ApJ...965L..14S}
}

@ARTICLE{2024MNRAS.529.2559V,
       author = {{van Velzen}, Sjoert and {Stein}, Robert and {Gilfanov}, Marat and {Kowalski}, Marek and {Hayasaki}, Kimitake and {Reusch}, Simeon and {Yao}, Yuhan and {Garrappa}, Simone and {Franckowiak}, Anna and {Gezari}, Suvi and {Nordin}, Jakob and {Fremling}, Christoffer and {Sharma}, Yashvi and {Yan}, Lin and {Kool}, Erik C. and {Stern}, Daniel and {Veres}, Patrik M. and {Sollerman}, Jesper and {Medvedev}, Pavel and {Sunyaev}, Rashid and {Bellm}, Eric C. and {Dekany}, Richard G. and {Duev}, Dimitri A. and {Graham}, Matthew J. and {Kasliwal}, Mansi M. and {Kulkarni}, Shrinivas R. and {Laher}, Russ R. and {Riddle}, Reed L. and {Rusholme}, Ben},
        title = "{Establishing accretion flares from supermassive black holes as a source of high-energy neutrinos}",
      journal = {\mnras},
         year = 2024,
        month = apr,
       volume = {529},
       number = {3},
        pages = {2559-2576},
          doi = {10.1093/mnras/stae610},
archivePrefix = {arXiv},
       eprint = {2111.09391},
 primaryClass = {astro-ph.HE},
       adsurl = {https://ui.adsabs.harvard.edu/abs/2024MNRAS.529.2559V}
}

@ARTICLE{2019ApJ...873..111I,
       author = {{Ivezi{\'c}}, {\v{Z}}eljko and {Kahn}, Steven M. and {Tyson}, J. Anthony and {Abel}, Bob and {Acosta}, Emily and {Allsman}, Robyn and {Alonso}, David and {AlSayyad}, Yusra and {Anderson}, Scott F. and {Andrew}, John and {Angel}, James Roger P. and {Angeli}, George Z. and {Ansari}, Reza and {Antilogus}, Pierre and {Araujo}, Constanza and {Armstrong}, Robert and {Arndt}, Kirk T. and {Astier}, Pierre and {Aubourg}, {\'E}ric and {Auza}, Nicole and {Axelrod}, Tim S. and {Bard}, Deborah J. and {Barr}, Jeff D. and {Barrau}, Aurelian and {Bartlett}, James G. and {Bauer}, Amanda E. and {Bauman}, Brian J. and {Baumont}, Sylvain and {Bechtol}, Ellen and {Bechtol}, Keith and {Becker}, Andrew C. and {Becla}, Jacek and {Beldica}, Cristina and {Bellavia}, Steve and {Bianco}, Federica B. and {Biswas}, Rahul and {Blanc}, Guillaume and {Blazek}, Jonathan and {Blandford}, Roger D. and {Bloom}, Josh S. and {Bogart}, Joanne and {Bond}, Tim W. and {Booth}, Michael T. and {Borgland}, Anders W. and {Borne}, Kirk and {Bosch}, James F. and {Boutigny}, Dominique and {Brackett}, Craig A. and {Bradshaw}, Andrew and {Brandt}, William Nielsen and {Brown}, Michael E. and {Bullock}, James S. and {Burchat}, Patricia and {Burke}, David L. and {Cagnoli}, Gianpietro and {Calabrese}, Daniel and {Callahan}, Shawn and {Callen}, Alice L. and {Carlin}, Jeffrey L. and {Carlson}, Erin L. and {Chandrasekharan}, Srinivasan and {Charles-Emerson}, Glenaver and {Chesley}, Steve and {Cheu}, Elliott C. and {Chiang}, Hsin-Fang and {Chiang}, James and {Chirino}, Carol and {Chow}, Derek and {Ciardi}, David R. and {Claver}, Charles F. and {Cohen-Tanugi}, Johann and {Cockrum}, Joseph J. and {Coles}, Rebecca and {Connolly}, Andrew J. and {Cook}, Kem H. and {Cooray}, Asantha and {Covey}, Kevin R. and {Cribbs}, Chris and {Cui}, Wei and {Cutri}, Roc and {Daly}, Philip N. and {Daniel}, Scott F. and {Daruich}, Felipe and {Daubard}, Guillaume and {Daues}, Greg and {Dawson}, William and {Delgado}, Francisco and {Dellapenna}, Alfred and {de Peyster}, Robert and {de Val-Borro}, Miguel and {Digel}, Seth W. and {Doherty}, Peter and {Dubois}, Richard and {Dubois-Felsmann}, Gregory P. and {Durech}, Josef and {Economou}, Frossie and {Eifler}, Tim and {Eracleous}, Michael and {Emmons}, Benjamin L. and {Fausti Neto}, Angelo and {Ferguson}, Henry and {Figueroa}, Enrique and {Fisher-Levine}, Merlin and {Focke}, Warren and {Foss}, Michael D. and {Frank}, James and {Freemon}, Michael D. and {Gangler}, Emmanuel and {Gawiser}, Eric and {Geary}, John C. and {Gee}, Perry and {Geha}, Marla and {Gessner}, Charles J.~B. and {Gibson}, Robert R. and {Gilmore}, D. Kirk and {Glanzman}, Thomas and {Glick}, William and {Goldina}, Tatiana and {Goldstein}, Daniel A. and {Goodenow}, Iain and {Graham}, Melissa L. and {Gressler}, William J. and {Gris}, Philippe and {Guy}, Leanne P. and {Guyonnet}, Augustin and {Haller}, Gunther and {Harris}, Ron and {Hascall}, Patrick A. and {Haupt}, Justine and {Hernandez}, Fabio and {Herrmann}, Sven and {Hileman}, Edward and {Hoblitt}, Joshua and {Hodgson}, John A. and {Hogan}, Craig and {Howard}, James D. and {Huang}, Dajun and {Huffer}, Michael E. and {Ingraham}, Patrick and {Innes}, Walter R. and {Jacoby}, Suzanne H. and {Jain}, Bhuvnesh and {Jammes}, Fabrice and {Jee}, M. James and {Jenness}, Tim and {Jernigan}, Garrett and {Jevremovi{\'c}}, Darko and {Johns}, Kenneth and {Johnson}, Anthony S. and {Johnson}, Margaret W.~G. and {Jones}, R. Lynne and {Juramy-Gilles}, Claire and {Juri{\'c}}, Mario and {Kalirai}, Jason S. and {Kallivayalil}, Nitya J. and {Kalmbach}, Bryce and {Kantor}, Jeffrey P. and {Karst}, Pierre and {Kasliwal}, Mansi M. and {Kelly}, Heather and {Kessler}, Richard and {Kinnison}, Veronica and {Kirkby}, David and {Knox}, Lloyd and {Kotov}, Ivan V. and {Krabbendam}, Victor L. and {Krughoff}, K. Simon and {Kub{\'a}nek}, Petr and {Kuczewski}, John and {Kulkarni}, Shri and {Ku}, John and {Kurita}, Nadine R. and {Lage}, Craig S. and {Lambert}, Ron and {Lange}, Travis and {Langton}, J. Brian and {Le Guillou}, Laurent and {Levine}, Deborah and {Liang}, Ming and {Lim}, Kian-Tat and {Lintott}, Chris J. and {Long}, Kevin E. and {Lopez}, Margaux and {Lotz}, Paul J. and {Lupton}, Robert H. and {Lust}, Nate B. and {MacArthur}, Lauren A. and {Mahabal}, Ashish and {Mandelbaum}, Rachel and {Markiewicz}, Thomas W. and {Marsh}, Darren S. and {Marshall}, Philip J. and {Marshall}, Stuart and {May}, Morgan and {McKercher}, Robert and {McQueen}, Michelle and {Meyers}, Joshua and {Migliore}, Myriam and {Miller}, Michelle and {Mills}, David J.},
        title = "{LSST: From Science Drivers to Reference Design and Anticipated Data Products}",
      journal = {\apj},
         year = 2019,
        month = mar,
       volume = {873},
       number = {2},
          eid = {111},
        pages = {111},
          doi = {10.3847/1538-4357/ab042c},
archivePrefix = {arXiv},
       eprint = {0805.2366},
 primaryClass = {astro-ph},
       adsurl = {https://ui.adsabs.harvard.edu/abs/2019ApJ...873..111I}
}

@ARTICLE{2015ApJ...811...52A,
       author = {{Aartsen}, M.~G. and {Abraham}, K. and {Ackermann}, M. and {Adams}, J. and {Aguilar}, J.~A. and {Ahlers}, M. and {Ahrens}, M. and {Altmann}, D. and {Anderson}, T. and {Archinger}, M. and {Arguelles}, C. and {Arlen}, T.~C. and {Auffenberg}, J. and {Bai}, X. and {Barwick}, S.~W. and {Baum}, V. and {Bay}, R. and {Beatty}, J.~J. and {Becker Tjus}, J. and {Becker}, K.-H. and {Beiser}, E. and {BenZvi}, S. and {Berghaus}, P. and {Berley}, D. and {Bernardini}, E. and {Bernhard}, A. and {Besson}, D.~Z. and {Binder}, G. and {Bindig}, D. and {Bissok}, M. and {Blaufuss}, E. and {Blumenthal}, J. and {Boersma}, D.~J. and {Bohm}, C. and {B{\"o}rner}, M. and {Bos}, F. and {Bose}, D. and {B{\"o}ser}, S. and {Botner}, O. and {Braun}, J. and {Brayeur}, L. and {Bretz}, H.-P. and {Brown}, A.~M. and {Buzinsky}, N. and {Casey}, J. and {Casier}, M. and {Cheung}, E. and {Chirkin}, D. and {Christov}, A. and {Christy}, B. and {Clark}, K. and {Classen}, L. and {Coenders}, S. and {Cowen}, D.~F. and {Cruz Silva}, A.~H. and {Daughhetee}, J. and {Davis}, J.~C. and {Day}, M. and {de Andr{\'e}}, J.~P.~A.~M. and {De Clercq}, C. and {Dembinski}, H. and {De Ridder}, S. and {Desiati}, P. and {de Vries}, K.~D. and {de Wasseige}, G. and {de With}, M. and {DeYoung}, T. and {D{\'\i}az-V{\'e}lez}, J.~C. and {Dumm}, J.~P. and {Dunkman}, M. and {Eagan}, R. and {Eberhardt}, B. and {Ehrhardt}, T. and {Eichmann}, B. and {Euler}, S. and {Evenson}, P.~A. and {Fadiran}, O. and {Fahey}, S. and {Fazely}, A.~R. and {Fedynitch}, A. and {Feintzeig}, J. and {Felde}, J. and {Filimonov}, K. and {Finley}, C. and {Fischer-Wasels}, T. and {Flis}, S. and {Fuchs}, T. and {Glagla}, M. and {Gaisser}, T.~K. and {Gaior}, R. and {Gallagher}, J. and {Gerhardt}, L. and {Ghorbani}, K. and {Gier}, D. and {Gladstone}, L. and {Gl{\"u}senkamp}, T. and {Goldschmidt}, A. and {Golup}, G. and {Gonzalez}, J.~G. and {G{\'o}ra}, D. and {Grant}, D. and {Gretskov}, P. and {Groh}, J.~C. and {Gross}, A. and {Ha}, C. and {Haack}, C. and {Haj Ismail}, A. and {Hallgren}, A. and {Halzen}, F. and {Hansmann}, B. and {Hanson}, K. and {Hebecker}, D. and {Heereman}, D. and {Helbing}, K. and {Hellauer}, R. and {Hellwig}, D. and {Hickford}, S. and {Hignight}, J. and {Hill}, G.~C. and {Hoffman}, K.~D. and {Hoffmann}, R. and {Holzapfe}, K. and {Homeier}, A. and {Hoshina}, K. and {Huang}, F. and {Huber}, M. and {Huelsnitz}, W. and {Hulth}, P.~O. and {Hultqvist}, K. and {In}, S. and {Ishihara}, A. and {Jacobi}, E. and {Japaridze}, G.~S. and {Jero}, K. and {Jurkovic}, M. and {Kaminsky}, B. and {Kappes}, A. and {Karg}, T. and {Karle}, A. and {Kauer}, M. and {Keivani}, A. and {Kelley}, J.~L. and {Kemp}, J. and {Kheirandish}, A. and {Kiryluk}, J. and {Kl{\"a}s}, J. and {Klein}, S.~R. and {Kohnen}, G. and {Koirala}, R. and {Kolanoski}, H. and {Konietz}, R. and {Koob}, A. and {K{\"o}pke}, L. and {Kopper}, C. and {Kopper}, S. and {Koskinen}, D.~J. and {Kowalski}, M. and {Krings}, K. and {Kroll}, G. and {Kroll}, M. and {Kunnen}, J. and {Kurahashi}, N. and {Kuwabara}, T. and {Labare}, M. and {Lanfranchi}, J.~L. and {Larson}, M.~J. and {Lesiak-Bzdak}, M. and {Leuermann}, M. and {Leuner}, J. and {L{\"u}nemann}, J. and {Madsen}, J. and {Maggi}, G. and {Mahn}, K.~B.~M. and {Maruyama}, R. and {Mase}, K. and {Matis}, H.~S. and {Maunu}, R. and {McNally}, F. and {Meagher}, K. and {Medici}, M. and {Meli}, A. and {Menne}, T. and {Merino}, G. and {Meures}, T. and {Miarecki}, S. and {Middell}, E. and {Middlemas}, E. and {Miller}, J. and {Mohrmann}, L. and {Montaruli}, T. and {Morse}, R. and {Nahnhauer}, R. and {Naumann}, U. and {Niederhausen}, H. and {Nowicki}, S.~C. and {Nygren}, D.~R. and {Obertacke}, A. and {Olivas}, A. and {Omairat}, A. and {O'Murchadha}, A.},
        title = "{The Detection of a Type IIn Supernova in Optical Follow-up Observations of IceCube Neutrino Events}",
      journal = {\apj},
         year = 2015,
        month = sep,
       volume = {811},
       number = {1},
          eid = {52},
        pages = {52},
          doi = {10.1088/0004-637X/811/1/52},
archivePrefix = {arXiv},
       eprint = {1506.03115},
 primaryClass = {astro-ph.HE},
       adsurl = {https://ui.adsabs.harvard.edu/abs/2015ApJ...811...52A}
}

@ARTICLE{2021ApJ...920...56F,
       author = {{Frederick}, Sara and {Gezari}, Suvi and {Graham}, Matthew J. and {Sollerman}, Jesper and {van Velzen}, Sjoert and {Perley}, Daniel A. and {Stern}, Daniel and {Ward}, Charlotte and {Hammerstein}, Erica and {Hung}, Tiara and et al.},
        title = "{A Family Tree of Optical Transients from Narrow-line Seyfert 1 Galaxies}",
      journal = {\apj},
         year = 2021,
        month = oct,
       volume = {920},
       number = {1},
          eid = {56},
        pages = {56},
          doi = {10.3847/1538-4357/ac110f},
archivePrefix = {arXiv},
       eprint = {2010.08554},
 primaryClass = {astro-ph.HE},
       adsurl = {https://ui.adsabs.harvard.edu/abs/2021ApJ...920...56F}
}

@ARTICLE{2011PhRvD..84d3003M,
       author = {{Murase}, Kohta and {Thompson}, Todd A. and {Lacki}, Brian C. and {Beacom}, John F.},
        title = "{New class of high-energy transients from crashes of supernova ejecta with massive circumstellar material shells}",
      journal = {\prd},
         year = 2011,
        month = aug,
       volume = {84},
       number = {4},
          eid = {043003},
        pages = {043003},
          doi = {10.1103/PhysRevD.84.043003},
archivePrefix = {arXiv},
       eprint = {1012.2834},
 primaryClass = {astro-ph.HE},
       adsurl = {https://ui.adsabs.harvard.edu/abs/2011PhRvD..84d3003M}
}

@ARTICLE{2014PhRvD..90j3005F,
       author = {{Fang}, Ke and {Kotera}, Kumiko and {Murase}, Kohta and {Olinto}, Angela V.},
        title = "{Testing the newborn pulsar origin of ultrahigh energy cosmic rays with EeV neutrinos}",
      journal = {\prd},
         year = 2014,
        month = nov,
       volume = {90},
       number = {10},
          eid = {103005},
        pages = {103005},
          doi = {10.1103/PhysRevD.90.103005},
archivePrefix = {arXiv},
       eprint = {1311.2044},
 primaryClass = {astro-ph.HE},
       adsurl = {https://ui.adsabs.harvard.edu/abs/2014PhRvD..90j3005F}
}

@ARTICLE{2017MNRAS.470.1881P,
       author = {{Petropoulou}, M. and {Coenders}, S. and {Vasilopoulos}, G. and {Kamble}, A. and {Sironi}, L.},
        title = "{Point-source and diffuse high-energy neutrino emission from Type IIn supernovae}",
      journal = {\mnras},
         year = 2017,
        month = sep,
       volume = {470},
       number = {2},
        pages = {1881-1893},
          doi = {10.1093/mnras/stx1251},
archivePrefix = {arXiv},
       eprint = {1705.06752},
 primaryClass = {astro-ph.HE},
       adsurl = {https://ui.adsabs.harvard.edu/abs/2017MNRAS.470.1881P}
}

@ARTICLE{2023ApJ...948...42W,
       author = {{Winter}, Walter and {Lunardini}, Cecilia},
        title = "{Interpretation of the Observed Neutrino Emission from Three Tidal Disruption Events}",
      journal = {\apj},
         year = 2023,
        month = may,
       volume = {948},
       number = {1},
          eid = {42},
        pages = {42},
          doi = {10.3847/1538-4357/acbe9e},
archivePrefix = {arXiv},
       eprint = {2205.11538},
 primaryClass = {astro-ph.HE},
       adsurl = {https://ui.adsabs.harvard.edu/abs/2023ApJ...948...42W}
}

@ARTICLE{2022MNRAS.514.4406W,
       author = {{Wu}, Han-Ji and {Mou}, Guobin and {Wang}, Kai and {Wang}, Wei and {Li}, Zhuo},
        title = "{Could TDE outflows produce the PeV neutrino events?}",
      journal = {\mnras},
         year = 2022,
        month = aug,
       volume = {514},
       number = {3},
        pages = {4406-4412},
          doi = {10.1093/mnras/stac1621},
archivePrefix = {arXiv},
       eprint = {2112.01748},
 primaryClass = {astro-ph.HE},
       adsurl = {https://ui.adsabs.harvard.edu/abs/2022MNRAS.514.4406W}
}

@INPROCEEDINGS{2016SPIE.9908E..1MT,
       author = {{Tamura}, Naoyuki and {Takato}, Naruhisa and {Shimono}, Atsushi and {Moritani}, Yuki and {Yabe}, Kiyoto and {Ishizuka}, Yuki and {Ueda}, Akitoshi and {Kamata}, Yukiko and {Aghazarian}, Hrand and {Arnouts}, St{\'e}phane and {Barban}, Gabriel and {Barkhouser}, Robert H. and {Borges}, Renato C. and {Braun}, David F. and {Carr}, Michael A. and {Chabaud}, Pierre-Yves and {Chang}, Yin-Chang and {Chen}, Hsin-Yo and {Chiba}, Masashi and {Chou}, Richard C.~Y. and {Chu}, You-Hua and {Cohen}, Judith and {de Almeida}, Rodrigo P. and {de Oliveira}, Antonio C. and {de Oliveira}, Ligia S. and {Dekany}, Richard G. and {Dohlen}, Kjetil and {dos Santos}, Jesulino B. and {dos Santos}, Leandro H. and {Ellis}, Richard and {Fabricius}, Maximilian and {Ferrand}, Didier and {Ferreira}, D{\'e}cio and {Golebiowski}, Mirek and {Greene}, Jenny E. and {Gross}, Johannes and {Gunn}, James E. and {Hammond}, Randolph and {Harding}, Albert and {Hart}, Murdock and {Heckman}, Timothy M. and {Hirata}, Christopher M. and {Ho}, Paul and {Hope}, Stephen C. and {Hovland}, Larry and {Hsu}, Shu-Fu and {Hu}, Yen-Shan and {Huang}, Ping-Jie and {Jaquet}, Marc and {Jing}, Yipeng and {Karr}, Jennifer and {Kimura}, Masahiko and {King}, Matthew E. and {Komatsu}, Eiichiro and {Le Brun}, Vincent and {Le F{\`e}vre}, Olivier and {Le Fur}, Arnaud and {Le Mignant}, David and {Ling}, Hung-Hsu and {Loomis}, Craig P. and {Lupton}, Robert H. and {Madec}, Fabrice and {Mao}, Peter and {Marrara}, Lucas S. and {Mendes de Oliveira}, Claudia and {Minowa}, Yosuke and {Morantz}, Chaz and {Murayama}, Hitoshi and {Murray}, Graham J. and {Ohyama}, Youichi and {Orndorff}, Joseph and {Pascal}, Sandrine and {Pereira}, Jefferson M. and {Reiley}, Daniel and {Reinecke}, Martin and {Ritter}, Andreas and {Roberts}, Mitsuko and {Schwochert}, Mark A. and {Seiffert}, Michael D. and {Smee}, Stephen A. and {Sodre}, Laerte and {Spergel}, David N. and {Steinkraus}, Aaron J. and {Strauss}, Michael A. and {Surace}, Christian and {Suto}, Yasushi and {Suzuki}, Nao and {Swinbank}, John and {Tait}, Philip J. and {Takada}, Masahiro and {Tamura}, Tomonori and {Tanaka}, Yoko and {Tresse}, Laurence and {Verducci}, Orlando and {Vibert}, Didier and {Vidal}, Clement and {Wang}, Shiang-Yu and {Wen}, Chih-Yi and {Yan}, Chi-Hung and {Yasuda}, Naoki},
        title = "{Prime Focus Spectrograph (PFS) for the Subaru telescope: overview, recent progress, and future perspectives}",
    booktitle = {Ground-based and Airborne Instrumentation for Astronomy VI},
         year = 2016,
       editor = {{Evans}, Christopher J. and {Simard}, Luc and {Takami}, Hideki},
       series = {Society of Photo-Optical Instrumentation Engineers (SPIE) Conference Series},
       volume = {9908},
        month = aug,
          eid = {99081M},
        pages = {99081M},
          doi = {10.1117/12.2232103},
archivePrefix = {arXiv},
       eprint = {1608.01075},
 primaryClass = {astro-ph.IM},
       adsurl = {https://ui.adsabs.harvard.edu/abs/2016SPIE.9908E..1MT}
}

@ARTICLE{2024EPJC...84..885K,
       author = {{KM3NeT Collaboration} and {Aiello}, S. and {Albert}, A. and {Alshamsi}, M. and {Alves Garre}, S. and {Aly}, Z. and {Ambrosone}, A. and {Ameli}, F. and {Andre}, M. and {Androutsou}, E. and et al.},
        title = "{Astronomy potential of KM3NeT/ARCA}",
      journal = {European Physical Journal C},
         year = 2024,
        month = sep,
       volume = {84},
       number = {9},
          eid = {885},
        pages = {885},
          doi = {10.1140/epjc/s10052-024-13137-2},
archivePrefix = {arXiv},
       eprint = {2402.08363},
 primaryClass = {astro-ph.HE},
       adsurl = {https://ui.adsabs.harvard.edu/abs/2024EPJC...84..885K}
}

@ARTICLE{2020NatAs...4..913A,
       author = {{Agostini}, Matteo and {B{\"o}hmer}, Michael and {Bosma}, Jeff and {Clark}, Kenneth and {Danninger}, Matthias and {Fruck}, Christian and {Gernh{\"a}user}, Roman and {G{\"a}rtner}, Andreas and {Grant}, Darren and {Henningsen}, Felix and et al.},
        title = "{The Pacific Ocean Neutrino Experiment}",
      journal = {Nature Astronomy},
         year = 2020,
        month = sep,
       volume = {4},
        pages = {913-915},
          doi = {10.1038/s41550-020-1182-4},
archivePrefix = {arXiv},
       eprint = {2005.09493},
 primaryClass = {astro-ph.HE},
       adsurl = {https://ui.adsabs.harvard.edu/abs/2020NatAs...4..913A}
}

@ARTICLE{2023NatAs...7.1497Y,
       author = {{Ye}, Z.~P. and {Hu}, F. and {Tian}, W. and {Chang}, Q.~C. and {Chang}, Y.~L. and {Cheng}, Z.~S. and {Gao}, J. and {Ge}, T. and {Gong}, G.~H. and {Guo}, J. and et al.},
        title = "{A multi-cubic-kilometre neutrino telescope in the western Pacific Ocean}",
      journal = {Nature Astronomy},
         year = 2023,
        month = dec,
       volume = {7},
        pages = {1497-1505},
          doi = {10.1038/s41550-023-02087-6},
       adsurl = {https://ui.adsabs.harvard.edu/abs/2023NatAs...7.1497Y}
}

@ARTICLE{2025ApJ...981..159A,
       author = {{Abbasi}, R. and {Ackermann}, M. and {Adams}, J. and {Agarwalla}, S.~K. and {Aguilar}, J.~A. and {Ahlers}, M. and {Alameddine}, J.~M. and {Amin}, N.~M. and {Andeen}, K. and {Arg{\"u}elles}, C. and et al.},
        title = "{Search for Neutrino Doublets and Triplets Using 11.4 yr of IceCube Data}",
      journal = {\apj},
         year = 2025,
        month = mar,
       volume = {981},
       number = {2},
          eid = {159},
        pages = {159},
          doi = {10.3847/1538-4357/adb312},
archivePrefix = {arXiv},
       eprint = {2501.09276},
 primaryClass = {astro-ph.HE},
       adsurl = {https://ui.adsabs.harvard.edu/abs/2025ApJ...981..159A}
}

@ARTICLE{2019ApJ...883..125M,
       author = {{Morgan}, R. and {Bechtol}, K. and {Kessler}, R. and {Sako}, M. and {Herner}, K. and {Doctor}, Z. and {Scolnic}, D. and {Sevilla-Noarbe}, I. and {Franckowiak}, A. and {Neilson}, K.~N. and {Kowalski}, M. and {Palmese}, A. and {Swann}, E. and {Thomas}, B.~P. and {Vivas}, A.~K. and {Drlica-Wagner}, A. and {Garcia}, A. and {Brout}, D. and {Paz-Chinch{\'o}n}, F. and {Neilsen}, E. and {Diehl}, H.~T. and {Soares-Santos}, M. and {Abbott}, T.~M.~C. and {Avila}, S. and {Bertin}, E. and {Brooks}, D. and {Buckley-Geer}, E. and {Carnero Rosell}, A. and {Carrasco Kind}, M. and {Carretero}, J. and {Cawthon}, R. and {Costanzi}, M. and {De Vicente}, J. and {Desai}, S. and {Doel}, P. and {Flaugher}, B. and {Fosalba}, P. and {Frieman}, J. and {Garc{\'\i}a-Bellido}, J. and {Gaztanaga}, E. and {Gerdes}, D.~W. and {Gruen}, D. and {Gruendl}, R.~A. and {Gschwend}, J. and {Gutierrez}, G. and {Hollowood}, D.~L. and {Honscheid}, K. and {James}, D.~J. and {Kuropatkin}, N. and {Lima}, M. and {Maia}, M.~A.~G. and {Marshall}, J.~L. and {Menanteau}, F. and {Miller}, C.~J. and {Miquel}, R. and {Plazas}, A.~A. and {Sanchez}, E. and {Scarpine}, V. and {Schubnell}, M. and {Serrano}, S. and {Smith}, M. and {Sobreira}, F. and {Suchyta}, E. and {Swanson}, M.~E.~C. and {Tarle}, G. and {Vikram}, V. and {Walker}, A.~R. and {Weller}, J.},
        title = "{A DECam Search for Explosive Optical Transients Associated with IceCube Neutrino Alerts}",
      journal = {\apj},
         year = 2019,
        month = oct,
       volume = {883},
       number = {2},
          eid = {125},
        pages = {125},
          doi = {10.3847/1538-4357/ab3a45},
archivePrefix = {arXiv},
       eprint = {1907.07193},
 primaryClass = {astro-ph.HE},
       adsurl = {https://ui.adsabs.harvard.edu/abs/2019ApJ...883..125M}
}

@ARTICLE{2019A&A...626A.117P,
       author = {{Pan-STARRS Collaboration} and {Kankare}, E. and {Huber}, M. and {Smartt}, S.~J. and {Chambers}, K. and {Smith}, K.~W. and {McBrien}, O. and {Chen}, T.-W. and {Flewelling}, H. and {Lowe}, T. and {Magnier}, E. and {Schultz}, A. and {Waters}, C. and {Wainscoat}, R.~J. and {Willman}, M. and {Wright}, D. and {Young}, D. and {IceCube Collaboration} and {Aartsen}, M.~G. and {Ackermann}, M. and {Adams}, J. and {Aguilar}, J.~A. and {Ahlers}, M. and {Ahrens}, M. and {Alispach}, C. and {Altmann}, D. and {Andeen}, K. and {Anderson}, T. and {Ansseau}, I. and {Anton}, G. and {Arg{\"u}elles}, C. and {Auffenberg}, J. and {Axani}, S. and {Backes}, P. and {Bagherpour}, H. and {Bai}, X. and {Barbano}, A. and {Barwick}, S.~W. and {Baum}, V. and {Bay}, R. and {Beatty}, J.~J. and {Becker}, K.-H. and {Becker Tjus}, J. and {Benzvi}, S. and {Berley}, D. and {Bernardini}, E. and {Besson}, D.~Z. and {Binder}, G. and {Bindig}, D. and {Blaufuss}, E. and {Blot}, S. and {Bohm}, C. and {B{\"o}rner}, M. and {B{\"o}ser}, S. and {Botner}, O. and {Bourbeau}, E. and {Bourbeau}, J. and {Bradascio}, F. and {Braun}, J. and {Bretz}, H.-P. and {Bron}, S. and {Brostean-Kaiser}, J. and {Burgman}, A. and {Busse}, R.~S. and {Carver}, T. and {Chen}, C. and {Cheung}, E. and {Chirkin}, D. and {Clark}, K. and {Classen}, L. and {Collin}, G.~H. and {Conrad}, J.~M. and {Coppin}, P. and {Correa}, P. and {Cowen}, D.~F. and {Cross}, R. and {Dave}, P. and {de Andr{\'e}}, J.~P.~A.~M. and {de Clercq}, C. and {Delaunay}, J.~J. and {Dembinski}, H. and {Deoskar}, K. and {De Ridder}, S. and {Desiati}, P. and {de Vries}, K.~D. and {de Wasseige}, G. and {de With}, M. and {Deyoung}, T. and {D{\'\i}az-V{\'e}lez}, J.~C. and {Dujmovic}, H. and {Dunkman}, M. and {Dvorak}, E. and {Eberhardt}, B. and {Ehrhardt}, T. and {Eller}, P. and {Evenson}, P.~A. and {Fahey}, S. and {Fazely}, A.~R. and {Felde}, J. and {Filimonov}, K. and {Finley}, C. and {Franckowiak}, A. and {Friedman}, E. and {Fritz}, A. and {Gaisser}, T.~K. and {Gallagher}, J. and {Ganster}, E. and {Garrappa}, S. and {Gerhardt}, L. and {Ghorbani}, K. and {Glauch}, T. and {Gl{\"u}senkamp}, T. and {Goldschmidt}, A. and {Gonzalez}, J.~G. and {Grant}, D. and {Griffith}, Z. and {G{\"u}nd{\"u}z}, M. and {Haack}, C. and {Hallgren}, A. and {Halve}, L. and {Halzen}, F. and {Hanson}, K. and {Hebecker}, D. and {Heereman}, D. and {Helbing}, K. and {Hellauer}, R. and {Henningsen}, F. and {Hickford}, S. and {Hignight}, J. and {Hill}, G.~C. and {Hoffman}, K.~D. and {Hoffmann}, R. and {Hoinka}, T. and {Hokanson-Fasig}, B. and {Hoshina}, K. and {Huang}, F. and {Huber}, M. and {Hultqvist}, K. and {H{\"u}nnefeld}, M. and {Hussain}, R. and {in}, S. and {Iovine}, N. and {Ishihara}, A. and {Jacobi}, E. and {Japaridze}, G.~S. and {Jeong}, M. and {Jero}, K. and {Jones}, B.~J.~P. and {Kalaczynski}, P. and {Kang}, W. and {Kappes}, A. and {Kappesser}, D. and {Karg}, T. and {Karl}, M. and {Karle}, A. and {Katz}, U. and {Kauer}, M. and {Keivani}, A. and {Kelley}, J.~L. and {Kheirandish}, A. and {Kim}, J. and {Kintscher}, T. and {Kiryluk}, J. and {Kittler}, T. and {Klein}, S.~R. and {Koirala}, R. and {Kolanoski}, H. and {K{\"o}pke}, L. and {Kopper}, C. and {Kopper}, S. and {Koskinen}, D.~J. and {Kowalski}, M. and {Krings}, K. and {Kr{\"u}ckl}, G. and {Kulacz}, N. and {Kunwar}, S. and {Kurahashi}, N. and {Kyriacou}, A. and {Labare}, M. and {Lanfranchi}, J.~L. and {Larson}, M.~J. and {Lauber}, F. and {Lazar}, J.~P. and {Leonard}, K. and {Leuermann}, M. and {Liu}, Q.~R. and {Lohfink}, E. and {Lozano Mariscal}, C.~J. and {Lu}, L. and {Lucarelli}, F. and {L{\"u}nemann}, J. and {Luszczak}, W. and {Madsen}, J. and {Maggi}, G. and {Mahn}, K.~B.~M. and {Makino}, Y. and {Mallot}, K. and {Mancina}, S. and {Mari{\textcommabelow s}}, I.~C. and {Maruyama}, R.},
        title = "{Search for transient optical counterparts to high-energy IceCube neutrinos with Pan-STARRS1}",
      journal = {\aap},
         year = 2019,
        month = jun,
       volume = {626},
          eid = {A117},
        pages = {A117},
          doi = {10.1051/0004-6361/201935171},
archivePrefix = {arXiv},
       eprint = {1901.11080},
 primaryClass = {astro-ph.HE},
       adsurl = {https://ui.adsabs.harvard.edu/abs/2019A&A...626A.117P}
}

@ARTICLE{2017APh....92...30A,
       author = {{Aartsen}, M.~G. and {Ackermann}, M. and {Adams}, J. and {Aguilar}, J.~A. and {Ahlers}, M. and {Ahrens}, M. and {Altmann}, D. and {Andeen}, K. and {Anderson}, T. and {Ansseau}, I. and et al.},
        title = "{The IceCube realtime alert system}",
      journal = {Astroparticle Physics},
         year = 2017,
        month = jun,
       volume = {92},
        pages = {30-41},
          doi = {10.1016/j.astropartphys.2017.05.002},
archivePrefix = {arXiv},
       eprint = {1612.06028},
 primaryClass = {astro-ph.HE},
       adsurl = {https://ui.adsabs.harvard.edu/abs/2017APh....92...30A}
}

@ARTICLE{2021NatAs...5..510S,
       author = {{Stein}, Robert and {van Velzen}, Sjoert and {Kowalski}, Marek and {Franckowiak}, Anna and {Gezari}, Suvi and {Miller-Jones}, James C.~A. and {Frederick}, Sara and {Sfaradi}, Itai and {Bietenholz}, Michael F. and {Horesh}, Assaf and {Fender}, Rob and {Garrappa}, Simone and {Ahumada}, Tom{\'a}s and {Andreoni}, Igor and {Belicki}, Justin and {Bellm}, Eric C. and {B{\"o}ttcher}, Markus and {Brinnel}, Valery and {Burruss}, Rick and {Cenko}, S. Bradley and {Coughlin}, Michael W. and {Cunningham}, Virginia and {Drake}, Andrew and {Farrar}, Glennys R. and {Feeney}, Michael and {Foley}, Ryan J. and {Gal-Yam}, Avishay and {Golkhou}, V. Zach and {Goobar}, Ariel and {Graham}, Matthew J. and {Hammerstein}, Erica and {Helou}, George and {Hung}, Tiara and {Kasliwal}, Mansi M. and {Kilpatrick}, Charles D. and {Kong}, Albert K.~H. and {Kupfer}, Thomas and {Laher}, Russ R. and {Mahabal}, Ashish A. and {Masci}, Frank J. and {Necker}, Jannis and {Nordin}, Jakob and {Perley}, Daniel A. and {Rigault}, Mickael and {Reusch}, Simeon and {Rodriguez}, Hector and {Rojas-Bravo}, C{\'e}sar and {Rusholme}, Ben and {Shupe}, David L. and {Singer}, Leo P. and {Sollerman}, Jesper and {Soumagnac}, Maayane T. and {Stern}, Daniel and {Taggart}, Kirsty and {van Santen}, Jakob and {Ward}, Charlotte and {Woudt}, Patrick and {Yao}, Yuhan},
        title = "{A tidal disruption event coincident with a high-energy neutrino}",
      journal = {Nature Astronomy},
         year = 2021,
        month = feb,
       volume = {5},
        pages = {510-518},
          doi = {10.1038/s41550-020-01295-8},
archivePrefix = {arXiv},
       eprint = {2005.05340},
 primaryClass = {astro-ph.HE},
       adsurl = {https://ui.adsabs.harvard.edu/abs/2021NatAs...5..510S}
}

@ARTICLE{2022PhRvL.128v1101R,
       author = {{Reusch}, Simeon and {Stein}, Robert and {Kowalski}, Marek and {van Velzen}, Sjoert and {Franckowiak}, Anna and {Lunardini}, Cecilia and {Murase}, Kohta and {Winter}, Walter and {Miller-Jones}, James C.~A. and {Kasliwal}, Mansi M. and {Gilfanov}, Marat and {Garrappa}, Simone and {Paliya}, Vaidehi S. and {Ahumada}, Tom{\'a}s and {Anand}, Shreya and {Barbarino}, Cristina and {Bellm}, Eric C. and {Brinnel}, Val{\'e}ry and {Buson}, Sara and {Cenko}, S. Bradley and {Coughlin}, Michael W. and {De}, Kishalay and {Dekany}, Richard and {Frederick}, Sara and {Gal-Yam}, Avishay and {Gezari}, Suvi and {Giroletti}, Marcello and {Graham}, Matthew J. and {Karambelkar}, Viraj and {Kimura}, Shigeo S. and {Kong}, Albert K.~H. and {Kool}, Erik C. and {Laher}, Russ R. and {Medvedev}, Pavel and {Necker}, Jannis and {Nordin}, Jakob and {Perley}, Daniel A. and {Rigault}, Mickael and {Rusholme}, Ben and {Schulze}, Steve and {Schweyer}, Tassilo and {Singer}, Leo P. and {Sollerman}, Jesper and {Strotjohann}, Nora Linn and {Sunyaev}, Rashid and {van Santen}, Jakob and {Walters}, Richard and {Zhang}, B. Theodore and {Zimmerman}, Erez},
        title = "{Candidate Tidal Disruption Event AT2019fdr Coincident with a High-Energy Neutrino}",
      journal = {\prl},
         year = 2022,
        month = jun,
       volume = {128},
       number = {22},
          eid = {221101},
        pages = {221101},
          doi = {10.1103/PhysRevLett.128.221101},
archivePrefix = {arXiv},
       eprint = {2111.09390},
 primaryClass = {astro-ph.HE},
       adsurl = {https://ui.adsabs.harvard.edu/abs/2022PhRvL.128v1101R}
}

@ARTICLE{2023MNRAS.521.5046S,
       author = {{Stein}, Robert and {Reusch}, Simeon and {Franckowiak}, Anna and {Kowalski}, Marek and {Necker}, Jannis and {Weimann}, Sven and {Kasliwal}, Mansi M. and {Sollerman}, Jesper and {Ahumada}, Tomas and {Amaro Seoane}, Pau and {Anand}, Shreya and {Andreoni}, Igor and {Bellm}, Eric C. and {Bloom}, Joshua S. and {Coughlin}, Michael and {De}, Kishalay and {Fremling}, Christoffer and {Gezari}, Suvi and {Graham}, Matthew and {Groom}, Steven L. and {Helou}, George and {Kaplan}, David L. and {Karambelkar}, Viraj and {Kong}, Albert K.~H. and {Kool}, Erik C. and {Lincetto}, Massimiliano and {Mahabal}, Ashish A. and {Masci}, Frank J. and {Medford}, Michael S. and {Morgan}, Robert and {Nordin}, Jakob and {Rodriguez}, Hector and {Sharma}, Yashvi and {van Santen}, Jakob and {van Velzen}, Sjoert and {Yan}, Lin},
        title = "{Neutrino follow-up with the Zwicky transient facility: results from the first 24 campaigns}",
      journal = {\mnras},
         year = 2023,
        month = jun,
       volume = {521},
       number = {4},
        pages = {5046-5063},
          doi = {10.1093/mnras/stad767},
archivePrefix = {arXiv},
       eprint = {2203.17135},
 primaryClass = {astro-ph.HE},
       adsurl = {https://ui.adsabs.harvard.edu/abs/2023MNRAS.521.5046S}
}

@ARTICLE{2022MNRAS.516.2455N,
       author = {{Necker}, Jannis and {de Jaeger}, Thomas and {Stein}, Robert and {Franckowiak}, Anna and {Shappee}, Benjamin J. and {Kowalski}, Marek and {Kochanek}, Christopher S. and {Stanek}, Krzysztof Z. and {Beacom}, John F. and {Desai}, Dhvanil D. and {Neumann}, Kyle and {Jayasinghe}, Tharindu and {Holoien}, T.~W.-S. and {Thompson}, Todd A. and {Holmbo}, Simon},
        title = "{ASAS-SN follow-up of IceCube high-energy neutrino alerts}",
      journal = {\mnras},
         year = 2022,
        month = oct,
       volume = {516},
       number = {2},
        pages = {2455-2469},
          doi = {10.1093/mnras/stac2261},
archivePrefix = {arXiv},
       eprint = {2204.00500},
 primaryClass = {astro-ph.HE},
       adsurl = {https://ui.adsabs.harvard.edu/abs/2022MNRAS.516.2455N}
}

@ARTICLE{2026A&A...708A.223G,
       author = {{Garrappa}, S. and {Zimmerman}, E.~A. and {Wasserman}, T. and {Ofek}, E.~O. and {Gal-Yam}, A. and {Konno}, R. and {Chen}, P. and {Yaron}, O. and {Ben-Ami}, S. and {Copperwheat}, C.~M. and {Fainer}, S. and {Horowicz}, A. and {Humpe}, A. and {Mazzali}, P.~A. and {Polishook}, D. and {Segre}, E. and {Spitzer}, S.~A.},
        title = "{The Type IIn SN 2025cbj coincidence with the high-energy neutrino IceCube-250421A}",
      journal = {\aap},
         year = 2026,
        month = apr,
       volume = {708},
          eid = {A223},
        pages = {A223},
          doi = {10.1051/0004-6361/202558356},
archivePrefix = {arXiv},
       eprint = {2512.07936},
 primaryClass = {astro-ph.HE},
       adsurl = {https://ui.adsabs.harvard.edu/abs/2026A&A...708A.223G}
}

@ARTICLE{2025ApJ...993...23T,
       author = {{Toshikage}, Seiji and {Kimura}, Shigeo S. and {Shimizu}, Nobuhiro and {Tanaka}, Masaomi and {Yoshida}, Shigeru and {Iwakiri}, Wataru B. and {Morokuma}, Tomoki},
        title = "{The First Search for Optical Transient as a Counterpart of a Month-timescale IceCube Neutrino Multiplet Event}",
      journal = {\apj},
         year = 2025,
        month = nov,
       volume = {993},
       number = {1},
          eid = {23},
        pages = {23},
          doi = {10.3847/1538-4357/adfedf},
archivePrefix = {arXiv},
       eprint = {2504.04741},
 primaryClass = {astro-ph.HE},
       adsurl = {https://ui.adsabs.harvard.edu/abs/2025ApJ...993...23T}
}

@ARTICLE{2023ApJ...942....9H,
       author = {{Hammerstein}, Erica and {van Velzen}, Sjoert and {Gezari}, Suvi and {Cenko}, S. Bradley and {Yao}, Yuhan and {Ward}, Charlotte and {Frederick}, Sara and {Villanueva}, Natalia and {Somalwar}, Jean J. and {Graham}, Matthew J. and et al.},
        title = "{The Final Season Reimagined: 30 Tidal Disruption Events from the ZTF-I Survey}",
      journal = {\apj},
         year = 2023,
        month = jan,
       volume = {942},
       number = {1},
          eid = {9},
        pages = {9},
          doi = {10.3847/1538-4357/aca283},
archivePrefix = {arXiv},
       eprint = {2203.01461},
 primaryClass = {astro-ph.HE},
       adsurl = {https://ui.adsabs.harvard.edu/abs/2023ApJ...942....9H}
}

@ARTICLE{2007ApJ...665..265F,
       author = {{Faber}, S.~M. and {Willmer}, C.~N.~A. and {Wolf}, C. and {Koo}, D.~C. and {Weiner}, B.~J. and {Newman}, J.~A. and {Im}, M. and {Coil}, A.~L. and {Conroy}, C. and {Cooper}, M.~C. and et al.},
        title = "{Galaxy Luminosity Functions to z\raisebox{-0.5ex}\textasciitilde1 from DEEP2 and COMBO-17: Implications for Red Galaxy Formation}",
      journal = {\apj},
         year = 2007,
        month = aug,
       volume = {665},
       number = {1},
        pages = {265-294},
          doi = {10.1086/519294},
archivePrefix = {arXiv},
       eprint = {astro-ph/0506044},
 primaryClass = {astro-ph},
       adsurl = {https://ui.adsabs.harvard.edu/abs/2007ApJ...665..265F}
}

@ARTICLE{2007ApJ...663...81P,
       author = {{Polletta}, M. and {Tajer}, M. and {Maraschi}, L. and {Trinchieri}, G. and {Lonsdale}, C.~J. and {Chiappetti}, L. and {Andreon}, S. and {Pierre}, M. and {Le F{\`e}vre}, O. and {Zamorani}, G. and et al.},
        title = "{Spectral Energy Distributions of Hard X-Ray Selected Active Galactic Nuclei in the XMM-Newton Medium Deep Survey}",
      journal = {\apj},
         year = 2007,
        month = jul,
       volume = {663},
       number = {1},
        pages = {81-102},
          doi = {10.1086/518113},
archivePrefix = {arXiv},
       eprint = {astro-ph/0703255},
 primaryClass = {astro-ph},
       adsurl = {https://ui.adsabs.harvard.edu/abs/2007ApJ...663...81P}
}

@ARTICLE{2025A&A...695A.228N,
       author = {{Necker}, J. and {Graikou}, E. and {Kowalski}, M. and {Franckowiak}, A. and {Nordin}, J. and {Pernice}, T. and {van Velzen}, S. and {Veres}, P.~M.},
        title = "{Flaires: A comprehensive catalog of dust echo-like infrared flares}",
      journal = {\aap},
         year = 2025,
        month = mar,
       volume = {695},
          eid = {A228},
        pages = {A228},
          doi = {10.1051/0004-6361/202451340},
archivePrefix = {arXiv},
       eprint = {2407.01039},
 primaryClass = {astro-ph.HE},
       adsurl = {https://ui.adsabs.harvard.edu/abs/2025A&A...695A.228N}
}

@ARTICLE{2023A&A...674A..14R,
       author = {{Rimoldini}, Lorenzo and {Holl}, Berry and {Gavras}, Panagiotis and {Audard}, Marc and {De Ridder}, Joris and {Mowlavi}, Nami and {Nienartowicz}, Krzysztof and {Jevardat de Fombelle}, Gr{\'e}gory and {Lecoeur-Ta{\"\i}bi}, Isabelle and {Karbevska}, Lea and {Evans}, Dafydd W. and {{\'A}brah{\'a}m}, P{\'e}ter and {Carnerero}, Maria I. and {Clementini}, Gisella and {Distefano}, Elisa and {Garofalo}, Alessia and {Garc{\'\i}a-Lario}, Pedro and {Gomel}, Roy and {Klioner}, Sergei A. and {Kruszy{\'n}ska}, Katarzyna and {Lanzafame}, Alessandro C. and {Lebzelter}, Thomas and {Marton}, G{\'a}bor and {Mazeh}, Tsevi and {Molinaro}, Roberto and {Panahi}, Aviad and {Raiteri}, Claudia M. and {Ripepi}, Vincenzo and {Szabados}, L{\'a}szl{\'o} and {Teyssier}, David and {Trabucchi}, Michele and {Wyrzykowski}, {\L}ukasz and {Zucker}, Shay and {Eyer}, Laurent},
        title = "{Gaia Data Release 3. All-sky classification of 12.4 million variable sources into 25 classes}",
      journal = {\aap},
         year = 2023,
        month = jun,
       volume = {674},
          eid = {A14},
        pages = {A14},
          doi = {10.1051/0004-6361/202245591},
archivePrefix = {arXiv},
       eprint = {2211.17238},
 primaryClass = {astro-ph.GA},
       adsurl = {https://ui.adsabs.harvard.edu/abs/2023A&A...674A..14R}
}

@ARTICLE{1998PhRvD..57.3873F,
       author = {{Feldman}, Gary J. and {Cousins}, Robert D.},
        title = "{Unified approach to the classical statistical analysis of small signals}",
      journal = {\prd},
         year = 1998,
        month = apr,
       volume = {57},
       number = {7},
        pages = {3873-3889},
          doi = {10.1103/PhysRevD.57.3873},
archivePrefix = {arXiv},
       eprint = {physics/9711021},
 primaryClass = {physics.data-an},
       adsurl = {https://ui.adsabs.harvard.edu/abs/1998PhRvD..57.3873F}
}

@ARTICLE{2023ApJ...955L...6Y,
       author = {{Yao}, Yuhan and {Ravi}, Vikram and {Gezari}, Suvi and {van Velzen}, Sjoert and {Lu}, Wenbin and {Schulze}, Steve and {Somalwar}, Jean J. and {Kulkarni}, S.~R. and {Hammerstein}, Erica and {Nicholl}, Matt and et al.},
        title = "{Tidal Disruption Event Demographics with the Zwicky Transient Facility: Volumetric Rates, Luminosity Function, and Implications for the Local Black Hole Mass Function}",
      journal = {\apjl},
         year = 2023,
        month = sep,
       volume = {955},
       number = {1},
          eid = {L6},
        pages = {L6},
          doi = {10.3847/2041-8213/acf216},
archivePrefix = {arXiv},
       eprint = {2303.06523},
 primaryClass = {astro-ph.HE},
       adsurl = {https://ui.adsabs.harvard.edu/abs/2023ApJ...955L...6Y}
}

@ARTICLE{2022ApJ...937..108Y,
       author = {{Yoshida}, Shigeru and {Murase}, Kohta and {Tanaka}, Masaomi and {Shimizu}, Nobuhiro and {Ishihara}, Aya},
        title = "{Identifying High-energy Neutrino Transients by Neutrino Multiplet-triggered Follow-ups}",
      journal = {\apj},
         year = 2022,
        month = oct,
       volume = {937},
       number = {2},
          eid = {108},
        pages = {108},
          doi = {10.3847/1538-4357/ac8dfd},
archivePrefix = {arXiv},
       eprint = {2206.13719},
 primaryClass = {astro-ph.HE},
       adsurl = {https://ui.adsabs.harvard.edu/abs/2022ApJ...937..108Y}
}

@ARTICLE{2015ApJ...812...33S,
       author = {{Sun}, Hui and {Zhang}, Bing and {Li}, Zhuo},
        title = "{Extragalactic High-energy Transients: Event Rate Densities and Luminosity Functions}",
      journal = {\apj},
         year = 2015,
        month = oct,
       volume = {812},
       number = {1},
          eid = {33},
        pages = {33},
          doi = {10.1088/0004-637X/812/1/33},
archivePrefix = {arXiv},
       eprint = {1509.01592},
 primaryClass = {astro-ph.HE},
       adsurl = {https://ui.adsabs.harvard.edu/abs/2015ApJ...812...33S}
}

@ARTICLE{2018AJ....156..123A,
       author = {{Astropy Collaboration} and {Price-Whelan}, A.~M. and {Sip{\H{o}}cz}, B.~M. and {G{\"u}nther}, H.~M. and {Lim}, P.~L. and {Crawford}, S.~M. and {Conseil}, S. and {Shupe}, D.~L. and {Craig}, M.~W. and {Dencheva}, N. and {Ginsburg}, A. and {VanderPlas}, J.~T. and {Bradley}, L.~D. and {P{\'e}rez-Su{\'a}rez}, D. and {de Val-Borro}, M. and {Aldcroft}, T.~L. and {Cruz}, K.~L. and {Robitaille}, T.~P. and {Tollerud}, E.~J. and {Ardelean}, C. and {Babej}, T. and {Bach}, Y.~P. and {Bachetti}, M. and {Bakanov}, A.~V. and {Bamford}, S.~P. and {Barentsen}, G. and {Barmby}, P. and {Baumbach}, A. and {Berry}, K.~L. and {Biscani}, F. and {Boquien}, M. and {Bostroem}, K.~A. and {Bouma}, L.~G. and {Brammer}, G.~B. and {Bray}, E.~M. and {Breytenbach}, H. and {Buddelmeijer}, H. and {Burke}, D.~J. and {Calderone}, G. and {Cano Rodr{\'\i}guez}, J.~L. and {Cara}, M. and {Cardoso}, J.~V.~M. and {Cheedella}, S. and {Copin}, Y. and {Corrales}, L. and {Crichton}, D. and {D'Avella}, D. and {Deil}, C. and {Depagne}, {\'E}. and {Dietrich}, J.~P. and {Donath}, A. and {Droettboom}, M. and {Earl}, N. and {Erben}, T. and {Fabbro}, S. and {Ferreira}, L.~A. and {Finethy}, T. and {Fox}, R.~T. and {Garrison}, L.~H. and {Gibbons}, S.~L.~J. and {Goldstein}, D.~A. and {Gommers}, R. and {Greco}, J.~P. and {Greenfield}, P. and {Groener}, A.~M. and {Grollier}, F. and {Hagen}, A. and {Hirst}, P. and {Homeier}, D. and {Horton}, A.~J. and {Hosseinzadeh}, G. and {Hu}, L. and {Hunkeler}, J.~S. and {Ivezi{\'c}}, {\v{Z}}. and {Jain}, A. and {Jenness}, T. and {Kanarek}, G. and {Kendrew}, S. and {Kern}, N.~S. and {Kerzendorf}, W.~E. and {Khvalko}, A. and {King}, J. and {Kirkby}, D. and {Kulkarni}, A.~M. and {Kumar}, A. and {Lee}, A. and {Lenz}, D. and {Littlefair}, S.~P. and {Ma}, Z. and {Macleod}, D.~M. and {Mastropietro}, M. and {McCully}, C. and {Montagnac}, S. and {Morris}, B.~M. and {Mueller}, M. and {Mumford}, S.~J. and {Muna}, D. and {Murphy}, N.~A. and {Nelson}, S. and {Nguyen}, G.~H. and {Ninan}, J.~P. and {N{\"o}the}, M. and {Ogaz}, S. and {Oh}, S. and {Parejko}, J.~K. and {Parley}, N. and {Pascual}, S. and {Patil}, R. and {Patil}, A.~A. and {Plunkett}, A.~L. and {Prochaska}, J.~X. and {Rastogi}, T. and {Reddy Janga}, V. and {Sabater}, J. and {Sakurikar}, P. and {Seifert}, M. and {Sherbert}, L.~E. and {Sherwood-Taylor}, H. and {Shih}, A.~Y. and {Sick}, J. and {Silbiger}, M.~T. and {Singanamalla}, S. and {Singer}, L.~P. and {Sladen}, P.~H. and {Sooley}, K.~A. and {Sornarajah}, S. and {Streicher}, O. and {Teuben}, P. and {Thomas}, S.~W. and {Tremblay}, G.~R. and {Turner}, J.~E.~H. and {Terr{\'o}n}, V. and {van Kerkwijk}, M.~H. and {de la Vega}, A. and {Watkins}, L.~L. and {Weaver}, B.~A. and {Whitmore}, J.~B. and {Woillez}, J. and {Zabalza}, V. and {Astropy Contributors}},
        title = "{The Astropy Project: Building an Open-science Project and Status of the v2.0 Core Package}",
      journal = {\aj},
         year = 2018,
        month = sep,
       volume = {156},
       number = {3},
          eid = {123},
        pages = {123},
          doi = {10.3847/1538-3881/aabc4f},
archivePrefix = {arXiv},
       eprint = {1801.02634},
 primaryClass = {astro-ph.IM},
       adsurl = {https://ui.adsabs.harvard.edu/abs/2018AJ....156..123A}
}

@ARTICLE{2013A&A...558A..33A,
       author = {{Astropy Collaboration} and {Robitaille}, Thomas P. and
         {Tollerud}, Erik J. and {Greenfield}, Perry and {Droettboom}, Michael and
         {Bray}, Erik and {Aldcroft}, Tom and {Davis}, Matt and
         {Ginsburg}, Adam and {Price-Whelan}, Adrian M. and
         {Kerzendorf}, Wolfgang E. and {Conley}, Alexander and {Crighton}, Neil and
         {Barbary}, Kyle and {Muna}, Demitri and {Ferguson}, Henry and
         {Grollier}, Fr{\'e}d{\'e}ric and {Parikh}, Madhura M. and
         {Nair}, Prasanth H. and {Unther}, Hans M. and {Deil}, Christoph and
         {Woillez}, Julien and {Conseil}, Simon and {Kramer}, Roban and
         {Turner}, James E.~H. and {Singer}, Leo and {Fox}, Ryan and
         {Weaver}, Benjamin A. and {Zabalza}, Victor and {Edwards}, Zachary I. and
         {Azalee Bostroem}, K. and {Burke}, D.~J. and {Casey}, Andrew R. and
         {Crawford}, Steven M. and {Dencheva}, Nadia and {Ely}, Justin and
         {Jenness}, Tim and {Labrie}, Kathleen and {Lim}, Pey Lian and
         {Pierfederici}, Francesco and {Pontzen}, Andrew and {Ptak}, Andy and
         {Refsdal}, Brian and {Servillat}, Mathieu and {Streicher}, Ole},
        title = "{Astropy: A community Python package for astronomy}",
      journal = {\aap},
         year = "2013",
        month = "Oct",
       volume = {558},
          eid = {A33},
        pages = {A33},
          doi = {10.1051/0004-6361/201322068},
archivePrefix = {arXiv},
       eprint = {1307.6212},
 primaryClass = {astro-ph.IM},
       adsurl = {https://ui.adsabs.harvard.edu/abs/2013A&A...558A..33A}
}

@ARTICLE{aihara18ssp,
       author = {{Aihara}, Hiroaki and {Arimoto}, Nobuo and {Armstrong}, Robert and {Arnouts}, St{\'e}phane and {Bahcall}, Neta A. and {Bickerton}, Steven and {Bosch}, James and {Bundy}, Kevin and {Capak}, Peter L. and {Chan}, James H.~H. and {Chiba}, Masashi and {Coupon}, Jean and {Egami}, Eiichi and {Enoki}, Motohiro and {Finet}, Francois and {Fujimori}, Hiroki and {Fujimoto}, Seiji and {Furusawa}, Hisanori and {Furusawa}, Junko and {Goto}, Tomotsugu and {Goulding}, Andy and {Greco}, Johnny P. and {Greene}, Jenny E. and {Gunn}, James E. and {Hamana}, Takashi and {Harikane}, Yuichi and {Hashimoto}, Yasuhiro and {Hattori}, Takashi and {Hayashi}, Masao and {Hayashi}, Yusuke and {He{\l}miniak}, Krzysztof G. and {Higuchi}, Ryo and {Hikage}, Chiaki and {Ho}, Paul T.~P. and {Hsieh}, Bau-Ching and {Huang}, Kuiyun and {Huang}, Song and {Ikeda}, Hiroyuki and {Imanishi}, Masatoshi and {Inoue}, Akio K. and {Iwasawa}, Kazushi and {Iwata}, Ikuru and {Jaelani}, Anton T. and {Jian}, Hung-Yu and {Kamata}, Yukiko and {Karoji}, Hiroshi and {Kashikawa}, Nobunari and {Katayama}, Nobuhiko and {Kawanomoto}, Satoshi and {Kayo}, Issha and {Koda}, Jin and {Koike}, Michitaro and {Kojima}, Takashi and {Komiyama}, Yutaka and {Konno}, Akira and {Koshida}, Shintaro and {Koyama}, Yusei and {Kusakabe}, Haruka and {Leauthaud}, Alexie and {Lee}, Chien-Hsiu and {Lin}, Lihwai and {Lin}, Yen-Ting and {Lupton}, Robert H. and {Mandelbaum}, Rachel and {Matsuoka}, Yoshiki and {Medezinski}, Elinor and {Mineo}, Sogo and {Miyama}, Shoken and {Miyatake}, Hironao and {Miyazaki}, Satoshi and {Momose}, Rieko and {More}, Anupreeta and {More}, Surhud and {Moritani}, Yuki and {Moriya}, Takashi J. and {Morokuma}, Tomoki and {Mukae}, Shiro and {Murata}, Ryoma and {Murayama}, Hitoshi and {Nagao}, Tohru and {Nakata}, Fumiaki and {Niida}, Mana and {Niikura}, Hiroko and {Nishizawa}, Atsushi J. and {Obuchi}, Yoshiyuki and {Oguri}, Masamune and {Oishi}, Yukie and {Okabe}, Nobuhiro and {Okamoto}, Sakurako and {Okura}, Yuki and {Ono}, Yoshiaki and {Onodera}, Masato and {Onoue}, Masafusa and {Osato}, Ken and {Ouchi}, Masami and {Price}, Paul A. and {Pyo}, Tae-Soo and {Sako}, Masao and {Sawicki}, Marcin and {Shibuya}, Takatoshi and {Shimasaku}, Kazuhiro and {Shimono}, Atsushi and {Shirasaki}, Masato and {Silverman}, John D. and {Simet}, Melanie and {Speagle}, Joshua and {Spergel}, David N. and {Strauss}, Michael A. and {Sugahara}, Yuma and {Sugiyama}, Naoshi and {Suto}, Yasushi and {Suyu}, Sherry H. and {Suzuki}, Nao and {Tait}, Philip J. and {Takada}, Masahiro and {Takata}, Tadafumi and {Tamura}, Naoyuki and {Tanaka}, Manobu M. and {Tanaka}, Masaomi and {Tanaka}, Masayuki and {Tanaka}, Yoko and {Terai}, Tsuyoshi and {Terashima}, Yuichi and {Toba}, Yoshiki and {Tominaga}, Nozomu and {Toshikawa}, Jun and {Turner}, Edwin L. and {Uchida}, Tomohisa and {Uchiyama}, Hisakazu and {Umetsu}, Keiichi and {Uraguchi}, Fumihiro and {Urata}, Yuji and {Usuda}, Tomonori and {Utsumi}, Yousuke and {Wang}, Shiang-Yu and {Wang}, Wei-Hao and {Wong}, Kenneth C. and {Yabe}, Kiyoto and {Yamada}, Yoshihiko and {Yamanoi}, Hitomi and {Yasuda}, Naoki and {Yeh}, Sherry and {Yonehara}, Atsunori and {Yuma}, Suraphong},
        title = "{The Hyper Suprime-Cam SSP Survey: Overview and survey design}",
      journal = {\pasj},
         year = 2018,
        month = jan,
       volume = {70},
          eid = {S4},
        pages = {S4},
          doi = {10.1093/pasj/psx066},
archivePrefix = {arXiv},
       eprint = {1704.05858},
 primaryClass = {astro-ph.IM},
       adsurl = {https://ui.adsabs.harvard.edu/abs/2018PASJ...70S...4A}
}

@INPROCEEDINGS{miyazaki06,
   author = {{Miyazaki}, S. and {Komiyama}, Y. and {Nakaya}, H. and {Doi}, Y. and 
	{Furusawa}, H. and {Gillingham}, P. and {Kamata}, Y. and {Takeshi}, K. and 
	{Nariai}, K.},
    title = "{HyperSuprime: project overview}",
booktitle = {Society of Photo-Optical Instrumentation Engineers (SPIE) Conference Series},
     year = 2006,
   series = {Society of Photo-Optical Instrumentation Engineers (SPIE) Conference Series},
   volume = 6269,
    month = jul,
      doi = {10.1117/12.672739},
   adsurl = {http://ads.nao.ac.jp/abs/2006SPIE.6269E...9M}
}

@ARTICLE{ic23gcn,
       author = {{IceCube Collaboration}},
        title = "{IceCube-230724A - IceCube observation of a high-energy neutrino candidate track-like event}",
      journal = {GRB Coordinates Network},
         year = 2023,
        month = jul,
       volume = {34265},
        pages = {1},
       adsurl = {https://ui.adsabs.harvard.edu/abs/2023GCN.34265....1I}
}

@ARTICLE{ivezic08,
       author = {{Ivezic}, Z. and {Axelrod}, T. and {Brandt}, W.~N. and {Burke}, D.~L. and {Claver}, C.~F. and {Connolly}, A. and {Cook}, K.~H. and {Gee}, P. and {Gilmore}, D.~K. and {Jacoby}, S.~H. and {Jones}, R.~L. and {Kahn}, S.~M. and {Kantor}, J.~P. and {Krabbendam}, V.~V. and {Lupton}, R.~H. and {Monet}, D.~G. and {Pinto}, P.~A. and {Saha}, A. and {Schalk}, T.~L. and {Schneider}, D.~P. and {Strauss}, M.~A. and {Stubbs}, C.~W. and {Sweeney}, D. and {Szalay}, A. and {Thaler}, J.~J. and {Tyson}, J.~A. and {LSST Collaboration}},
        title = "{Large Synoptic Survey Telescope: From Science Drivers To Reference Design}",
      journal = {Serbian Astronomical Journal},
         year = 2008,
        month = jun,
       volume = {176},
        pages = {1-13},
          doi = {10.2298/SAJ0876001I},
       adsurl = {https://ui.adsabs.harvard.edu/abs/2008SerAJ.176....1I}
}

@INPROCEEDINGS{Axelrod10,
       author = {{Axelrod}, T. and {Kantor}, J. and {Lupton}, R.~H. and {Pierfederici}, F.},
        title = "{An open source application framework for astronomical imaging pipelines}",
    booktitle = {Software and Cyberinfrastructure for Astronomy},
         year = 2010,
       editor = {{Radziwill}, Nicole M. and {Bridger}, Alan},
       series = {Society of Photo-Optical Instrumentation Engineers (SPIE) Conference Series},
       volume = {7740},
        month = jul,
          eid = {774015},
        pages = {774015},
          doi = {10.1117/12.857297},
       adsurl = {https://ui.adsabs.harvard.edu/abs/2010SPIE.7740E..15A}
}

@ARTICLE{chambers16,
       author = {{Chambers}, K.~C. and {Magnier}, E.~A. and {Metcalfe}, N. and {Flewelling}, H.~A. and {Huber}, M.~E. and {Waters}, C.~Z. and {Denneau}, L. and {Draper}, P.~W. and {Farrow}, D. and {Finkbeiner}, D.~P. and {Holmberg}, C. and {Koppenhoefer}, J. and {Price}, P.~A. and {Rest}, A. and {Saglia}, R.~P. and {Schlafly}, E.~F. and {Smartt}, S.~J. and {Sweeney}, W. and {Wainscoat}, R.~J. and {Burgett}, W.~S. and {Chastel}, S. and {Grav}, T. and {Heasley}, J.~N. and {Hodapp}, K.~W. and {Jedicke}, R. and {Kaiser}, N. and {Kudritzki}, R. -P. and {Luppino}, G.~A. and {Lupton}, R.~H. and {Monet}, D.~G. and {Morgan}, J.~S. and {Onaka}, P.~M. and {Shiao}, B. and {Stubbs}, C.~W. and {Tonry}, J.~L. and {White}, R. and {Ba{\~n}ados}, E. and {Bell}, E.~F. and {Bender}, R. and {Bernard}, E.~J. and {Boegner}, M. and {Boffi}, F. and {Botticella}, M.~T. and {Calamida}, A. and {Casertano}, S. and {Chen}, W. -P. and {Chen}, X. and {Cole}, S. and {Deacon}, N. and {Frenk}, C. and {Fitzsimmons}, A. and {Gezari}, S. and {Gibbs}, V. and {Goessl}, C. and {Goggia}, T. and {Gourgue}, R. and {Goldman}, B. and {Grant}, P. and {Grebel}, E.~K. and {Hambly}, N.~C. and {Hasinger}, G. and {Heavens}, A.~F. and {Heckman}, T.~M. and {Henderson}, R. and {Henning}, T. and {Holman}, M. and {Hopp}, U. and {Ip}, W. -H. and {Isani}, S. and {Jackson}, M. and {Keyes}, C.~D. and {Koekemoer}, A.~M. and {Kotak}, R. and {Le}, D. and {Liska}, D. and {Long}, K.~S. and {Lucey}, J.~R. and {Liu}, M. and {Martin}, N.~F. and {Masci}, G. and {McLean}, B. and {Mindel}, E. and {Misra}, P. and {Morganson}, E. and {Murphy}, D.~N.~A. and {Obaika}, A. and {Narayan}, G. and {Nieto-Santisteban}, M.~A. and {Norberg}, P. and {Peacock}, J.~A. and {Pier}, E.~A. and {Postman}, M. and {Primak}, N. and {Rae}, C. and {Rai}, A. and {Riess}, A. and {Riffeser}, A. and {Rix}, H.~W. and {R{\"o}ser}, S. and {Russel}, R. and {Rutz}, L. and {Schilbach}, E. and {Schultz}, A.~S.~B. and {Scolnic}, D. and {Strolger}, L. and {Szalay}, A. and {Seitz}, S. and {Small}, E. and {Smith}, K.~W. and {Soderblom}, D.~R. and {Taylor}, P. and {Thomson}, R. and {Taylor}, A.~N. and {Thakar}, A.~R. and {Thiel}, J. and {Thilker}, D. and {Unger}, D. and {Urata}, Y. and {Valenti}, J. and {Wagner}, J. and {Walder}, T. and {Walter}, F. and {Watters}, S.~P. and {Werner}, S. and {Wood-Vasey}, W.~M. and {Wyse}, R.},
        title = "{The Pan-STARRS1 Surveys}",
      journal = {arXiv e-prints},
         year = 2016,
        month = dec,
          eid = {arXiv:1612.05560},
        pages = {arXiv:1612.05560},
          doi = {10.48550/arXiv.1612.05560},
archivePrefix = {arXiv},
       eprint = {1612.05560},
 primaryClass = {astro-ph.IM},
       adsurl = {https://ui.adsabs.harvard.edu/abs/2016arXiv161205560C}
}

@ARTICLE{alard98,
   author = {{Alard}, C. and {Lupton}, R.~H.},
    title = "{A Method for Optimal Image Subtraction}",
  journal = {\apj},
   eprint = {astro-ph/9712287},
     year = 1998,
    month = aug,
   volume = 503,
    pages = {325-331},
      doi = {10.1086/305984},
   adsurl = {http://adsabs.harvard.edu/abs/1998ApJ...503..325A}
}

@ARTICLE{alard00,
   author = {{Alard}, C.},
    title = "{Image subtraction using a space-varying kernel}",
  journal = {\aaps},
     year = 2000,
    month = jun,
   volume = 144,
    pages = {363-370},
      doi = {10.1051/aas:2000214},
   adsurl = {http://adsabs.harvard.edu/abs/2000A%26AS..144..363A}
}

@ARTICLE{ohgami21,
       author = {{Ohgami}, Takayuki and {Tominaga}, Nozomu and {Utsumi}, Yousuke and {Niino}, Yuu and {Tanaka}, Masaomi and {Banerjee}, Smaranika and {Hamasaki}, Ryo and {Yoshida}, Michitoshi and {Terai}, Tsuyoshi and {Takagi}, Yuhei and {Morokuma}, Tomoki and {Sasada}, Mahito and {Akitaya}, Hiroshi and {Yasuda}, Naoki and {Yanagisawa}, Kenshi and {Ohsawa}, Ryou},
        title = "{Optical follow-up observation for GW event S190510g using Subaru/Hyper Suprime-Cam}",
      journal = {\pasj},
         year = 2021,
        month = apr,
       volume = {73},
       number = {2},
        pages = {350-364},
          doi = {10.1093/pasj/psab002},
archivePrefix = {arXiv},
       eprint = {2101.00800},
 primaryClass = {astro-ph.HE},
       adsurl = {https://ui.adsabs.harvard.edu/abs/2021PASJ...73..350O}
}

@ARTICLE{ohgami23,
       author = {{Ohgami}, Takayuki and {Becerra Gonz{\'a}lez}, Josefa and {Tominaga}, Nozomu and {Morokuma}, Tomoki and {Utsumi}, Yousuke and {Niino}, Yuu and {Tanaka}, Masaomi and {Banerjee}, Smaranika and {Poidevin}, Fr{\'e}d{\'e}rick and {Acosta-Pulido}, Jose Antonio and {P{\'e}rez-Fournon}, Ismael and {Mu{\~n}oz-Darias}, Teo and {Akitaya}, Hiroshi and {Yanagisawa}, Kenshi and {Sasada}, Mahito and {Yoshida}, Michitoshi and {Simunovic}, Mirko and {Ohsawa}, Ryou and {Tanaka}, Ichi and {Terai}, Tsuyoshi and {Takagi}, Yuhei and {J-GEM Collaboration}},
        title = "{Follow-up Survey for the Binary Black Hole Merger GW200224\_222234 Using Subaru/HSC and GTC/OSIRIS}",
      journal = {\apj},
         year = 2023,
        month = apr,
       volume = {947},
       number = {1},
          eid = {9},
        pages = {9},
          doi = {10.3847/1538-4357/acbd42},
archivePrefix = {arXiv},
       eprint = {2302.09269},
 primaryClass = {astro-ph.HE},
       adsurl = {https://ui.adsabs.harvard.edu/abs/2023ApJ...947....9O}
}

@ARTICLE{IceCube2022Sci,
       author = {{IceCube Collaboration} and {Abbasi}, R. and {Ackermann}, M. and {Adams}, J. and {Aguilar}, J.~A. and {Ahlers}, M. and {Ahrens}, M. and {Alameddine}, J.~M. and {Alispach}, C. and {Alves}, A.~A., Jr. and et al.},
        title = "{Evidence for neutrino emission from the nearby active galaxy NGC 1068}",
      journal = {Science},
         year = 2022,
        month = nov,
       volume = {378},
       number = {6619},
        pages = {538-543},
          doi = {10.1126/science.abg3395},
archivePrefix = {arXiv},
       eprint = {2211.09972},
 primaryClass = {astro-ph.HE},
       adsurl = {https://ui.adsabs.harvard.edu/abs/2022Sci...378..538I}
}

@article{IceCube2018blazar1,
      author         = "IceCube-Collaboration and Fermi-LAT and MAGIC and AGILE and ASAS-SN and HAWC and H.E.S.S and INTEGRAL and Kanata and Kiso and Kapteyn and Liverpool Telescope and Subaru and Swift/NuSTAR and VERITAS and VLA/17B-403 teams and others",
      collaboration  = "IceCube Collaboration",
      title          = "{Multimessenger observations of a flaring blazar coincident with high-energy neutrino IceCube-170922A}",
      journal        = "Science",
      volume         = "361",
      year           = "2018",
      pages          = "146",
      eprint         = "1807.08816",
      archivePrefix  = "arXiv",
      primaryClass   = "astro-ph.HE",
      SLACcitation   = "%%CITATION = ARXIV:1710.01191;%%"
}

@article{IceCube2018blazar2,
      author         = "IceCube-Collaboration",
      collaboration  = "IceCube Collaboration",
      title          = "{Neutrino emission from the direction of the blazar TXS 0506+056 prior to the IceCube-170922A alert}",
      journal        = "Science",
      volume         = "361",
      year           = "2018",
      pages          = "147",
      eprint         = "1807.08794",
      archivePrefix  = "arXiv",
      primaryClass   = "astro-ph.HE",
      SLACcitation   = "%%CITATION = ARXIV:1710.01191;%%"
}

@ARTICLE{IceCube2013PRL,
   author = {{Aartsen}, M.~G. and {Abbasi}, R. and {Abdou}, Y. and {Ackermann}, M. and 
	{Adams}, J. and {Aguilar}, J.~A. and {Ahlers}, M. and {Altmann}, D. and 
	{Auffenberg}, J. and {Bai}, X. and et al.},
    title = "{First Observation of PeV-Energy Neutrinos with IceCube}",
  journal = {Physical Review Letters},
archivePrefix = "arXiv",
   eprint = {1304.5356},
 primaryClass = "astro-ph.HE",
     year = 2013,
    month = jul,
   volume = 111,
   number = 2,
      eid = {021103},
    pages = {021103},
      doi = {10.1103/PhysRevLett.111.021103},
   adsurl = {http://adsabs.harvard.edu/abs/2013PhRvL.111b1103A}
}

@ARTICLE{IceCube2013Sci,
   author = {{IceCube Collaboration}},
    title = "{Evidence for High-Energy Extraterrestrial Neutrinos at the IceCube Detector}",
  journal = {Science},
archivePrefix = "arXiv",
   eprint = {1311.5238},
 primaryClass = "astro-ph.HE",
     year = 2013,
    month = nov,
   volume = 342,
      eid = {1242856},
    pages = {1242856},
      doi = {10.1126/science.1242856},
   adsurl = {http://adsabs.harvard.edu/abs/2013Sci...342E...1I}
}

@article{Murase:2020lnu,
    author = "Murase, Kohta and Kimura, Shigeo S. and Zhang, B. Theodore and Oikonomou, Foteini and Petropoulou, Maria",
    title = "{High-Energy Neutrino and Gamma-Ray Emission from Tidal Disruption Events}",
    eprint = "2005.08937",
    archivePrefix = "arXiv",
    primaryClass = "astro-ph.HE",
    doi = "10.3847/1538-4357/abb3c0",
    journal = "Astrophys. J.",
    volume = "902",
    number = "2",
    pages = "108",
    year = "2020"
}

@ARTICLE{hayasaki2019ApJ,
       author = {{Hayasaki}, Kimitake and {Yamazaki}, Ryo},
        title = "{Neutrino Emissions from Tidal Disruption Remnants}",
      journal = {\apj},
         year = "2019",
        month = "Dec",
       volume = {886},
       number = {2},
          eid = {114},
        pages = {114},
          doi = {10.3847/1538-4357/ab44ca},
       adsurl = {https://ui.adsabs.harvard.edu/abs/2019ApJ...886..114H}
}

@ARTICLE{Liu2020PhRvD,
       author = {{Liu}, Ruo-Yu and {Xi}, Shao-Qiang and {Wang}, Xiang-Yu},
        title = "{Neutrino emission from an off-axis jet driven by the tidal disruption event AT2019dsg}",
      journal = {\prd},
         year = 2020,
        month = oct,
       volume = {102},
       number = {8},
          eid = {083028},
        pages = {083028},
          doi = {10.1103/PhysRevD.102.083028},
archivePrefix = {arXiv},
       eprint = {2011.03773},
 primaryClass = {astro-ph.HE},
       adsurl = {https://ui.adsabs.harvard.edu/abs/2020PhRvD.102h3028L}
}

@ARTICLE{murase2018PhRvD,
       author = {{Murase}, Kohta},
        title = "{New prospects for detecting high-energy neutrinos from nearby supernovae}",
      journal = {\prd},
         year = "2018",
        month = "Apr",
       volume = {97},
       number = {8},
          eid = {081301},
        pages = {081301},
          doi = {10.1103/PhysRevD.97.081301},
archivePrefix = {arXiv},
       eprint = {1705.04750},
 primaryClass = {astro-ph.HE},
       adsurl = {https://ui.adsabs.harvard.edu/abs/2018PhRvD..97h1301M}
}
\bibliographystyle{aasjournal}



\end{document}